\documentclass{article}
\usepackage{psfig}
\usepackage{amsmath}
\usepackage[dvips]{graphicx}
\begin{document}

\title{Simple models of protein folding and of non--conventional drug design}
\large
\author{R.A. Broglia$^{1,2,3}$ G. Tiana$^{1,2}$\\
$^1$Department of Physics, University of Milano,\\
via Celoria 16, 20133 Milano, Italy,\\
$^2$INFN, Sez. di Milano, Milano, Italy,\\
$^3$The Niels Bohr Institute, University of Copenhagen,\\
Bledgamsvej 17, 2100 Copenhagen, Denmark}
\date{\today}

\maketitle
\begin{abstract}
While all the information required for the folding of a protein is contained in its amino acid sequence, one has not yet learned how to extract this information to predict the three--dimensional, biologically active, native conformation of a protein whose sequence is known. Using insight obtained from simple model simulations of the folding of proteins, in particular of the fact that this phenomenon is essentially controlled by conserved (native) contacts among (few) strongly interacting ("hot"), as a rule hydrophobic, amino acids, which also stabilize local elementary structures (LES, hidden, incipient secondary structures like $\alpha$--helices and $\beta$--sheets) formed early in the folding process and leading to the postcritical folding nucleus (i.e., the minimum set of native contacts which bring the system pass beyond the highest free--energy barrier found in the whole folding process) it is possible to work out a succesful strategy for reading the native structure of designed proteins from the knowledge of only their amino acid sequence and of the contact energies among the amino acids. Because LES have undergone millions of years of evolution to selectively dock to their complementary structures, small peptides made out of the same amino acids as the LES are expected to selectively attach to the newly expressed (unfolded) protein and inhibit its folding, or to the native (fluctuating) native conformation and denaturate it. These peptides, or their mimetic molecules, can thus be used as effective non--conventional drugs to those already existing (and directed at neutralizing the active site of enzymes), displaying the advantage of not suffering from the uprise of resistance. 
\end{abstract}

\parindent=0.5cm

\section{Introduction}
The sequencing of the human genome \cite{shoemaker,venter}, that is the identification of the way the thousands of milions of basis follow each other in the human DNA, provides   information on the sequence of amino acids forming each of the tens of thousands of proteins which build our cells and catalize the chemical reaction which make them function. Precious as this knowledge is, in terms of gene identification and thus eventually for the development of therapies against inherited diseases, this sequencing will find its real meaning when the linear sequence of the basis can be set in relation with the three dimensional, biologically active, native structure of the protein it codes for. 

\begin{figure}
\centerline{\psfig{file=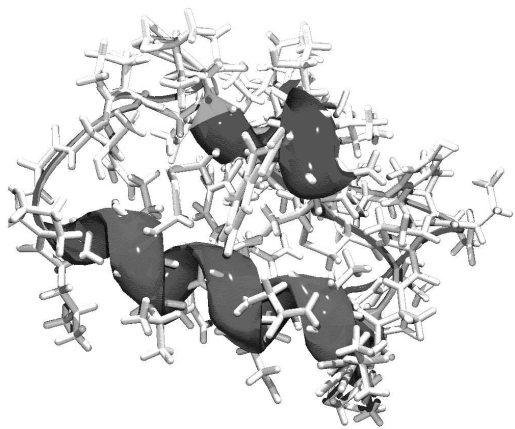,height=8cm}}
\caption{The atomic structure of a small protein, Chemotripsin Inhibitor 2. The dark grey curve highlights the chain structure.}
\label{protein}
\end{figure}

In a very real sense, the secret of life, if it exists, is to be found at this level of physical and chemical organization. In particular, in the tight correspondence  existing between the one dimensional (1D) structure of a protein (sequence of amino acids) and the native structure onto which it folds (3D structure, cf. Fig. \ref{protein}), once produced by the ribosome using the DNA (or better the mRNA) blueprint, in typical times which range from microseconds to seconds.  This is the protein folding problem, one of the great unsolved problems of science \cite{fersht}.  An even more elusive goal is the prediction of the catalytic activity of an enzyme from its amino acid sequence. 

There are two reasons why the protein folding problem is so important.  First, DNA sequencing is relatively quick, and vast quantities of amino acid sequences data have become available through international efforts.  The acquisition of three-dimensional data is still slow and is limited to proteins that either crystallize in a suitable form or are sufficiently small to be solved by NMR in solution \cite{service}.  Algorithms are thus required to translate the linear information into spatial information.  Second,one is now able to synthetize novel proteins by way of their genes, and so the production of new enzymes  with specified catalytic activities is a challenging prospect. Producing such new enzymes requires, at least, three underpinning and interrelated abilities: (1) the ability to design a novel fold of the enzyme or to predict the most stable and kinetically accessible conformation of an already existing protein if one is going to use a wild type conformation as a template; (2) the ability to design on this fold a binding site for the substrate and (3) the ability to build the catalytic site which is the ultimate responsible for the biological activity of the enzyme.  Each of these three prerequisites is beyond the current capabilities of theory. 

To appreciate some of the difficulties facing theoreticians, we must consider the physical nature of protein folding. A denatured protein makes many interactions with solvent water. As the protein folds up, it exchanges those noncovalent interactions with others that it makes within itself: its hydrophobic side chains (which cannot build any hydrogen-bond with water) tend to pack with one another, and many of its hydrogen bonds donors and acceptors pair with each other, especially those in the polypeptide backbone that form the hydrogen-bonded networks in helixes and sheets. Each interaction energy is small, but because of their large numbers, the total free energies in the native and denatured states are large, being some thousands of kilocalories per mole, depending on the size of the protein.  Yet proteins are only marginally stable, their free energies of unfolding ranging from 5 to 15 kcal/mol.  This tiny amount of energy is the difference between the free energies of the protein in its native and in its denatured states, and these states are almost ba\-lan\-ced in energy. Thus, whether a protein folds depends crucially on a balance between two large numbers, each of which is very    difficult to calculate with precision. To predict the stability of a protein, we have    to calculate not only the interaction energy between any two atoms within a protein, but also this energy relative to the interactions that the individual atoms make with water  in the denatured state and the entropic cost of folding, which represents the largest contribution to the unfolded free energy. Current potential functions are not sufficiently accurate for this purpose. But we can use protein engineering (that is introduce point mutations in the amino acid sequence) and other experimental procedures to make changes in existing proteins. Protein engineering experiments have provided a    practical route into determining quantitatively the factors that govern stability. 

It is unlikely that there is a single mechanism for the folding of all proteins. As biologists know so well, proteins vary so much in structure, size and properties that there are bound to be a number of important exceptions to any general scenario \cite{fersht}. Further, evolution of a specific function may be at the expense of stability or optimization of folding rate. In any case, the basic mechanism of small globular, single domain proteins all point to a single scheme, variations of which could describe a large number of folding pathways.  

Solving how a protein folds up from its denaturated state to its native structure poses an intellectual challenge that is far more complex than solving classical chemical mechanisms.  In simple chemical mechanisms, there are usually changes in just a small number of bonds   as a reaction progresses. Chemical bonds are often strong, and so stable covalent intermediates can sometimes be isolated and characterized. Often, the rules for analyzing mechanisms can be applied simply and rigorously. In protein folding, on the other hand,  the whole molecule changes in structure. Thousands of weak noncovalent
interactions are  made or broken, and it is very difficult to trap intermediates because of their unstable nature. An astronomical number of conformations must be considered. In particular, the   denaturated state is particularly difficult to analyze because it is an ensemble of many ill-defined rapidly fluctuating conformations. But the chemists basic strategy for analyzing the pathways of protein folding must still be the same as that for analyzing a simple chemical mechanism: characterize all the stable and metastable states on (and off) the reaction pathways and the transition states that link them. This knowledge will provide the basis for calculations and simulations, as well as for developing simple, but not oversimplified, models.

There is an arsenal of physical theoretical methods for the analysis of folding, ranging from precise atomic analysis by molecular dynamics simulations, through models involving simplified polymer chains, to completely abstract procedures \cite{varenna}. The most precise procedure of molecular dynamics simulation applies readily to unfolding   and can, in principle, take a (small) protein to the denaturated state in solution. Molecular dynamics simulations have not, as yet, been applied to the folding of the plethora of possible conformations. In comparison, the more abstract procedures allow simple models for folding to be analyzed directly, and they start from a state prior to the partly collapsed state  in solution. These procedures tend to stress general principles. All the methods can give insights into the principles of folding that can be tested against experiment, make predictions, and fill the entire energy landscape.

Understanding the folding of proteins is also likely to be momentous for the developement of both conventional \cite{smith} and non conventional drugs \cite{drug}.  Conventional drugs work by inhibiting the enzymatic activity of specific proteins by capping its active site. Pharmaceutical companies search for these drugs by a simple trial--and--error method, testing hundreds of thousands of molecules on the enzymatic target, selecting those which display the best inhibitory ability. Only recently some attempt is being made to go beyond such brute force method, by calculating theoretically the affinity of the molecule to the active site of the enzyme \cite{sh_drug}. Simple models which employ statistically derived potentials have been proved useful to perform such calculations in simple cases, but fail grievously when the quantum mechanical properties of the molecules enter the game, such as in the case of metallo--proteins.

The use of {\it ab initio} methods, like e.g. the density functional theory (DFT), to treat the properties of the "problematic" atoms, combined with a classical description of the surrounding molecules, should allow one to accurately calculate the binding energy of the ligand to the enzyme also for systems of realistic size (cf. e.g. \cite{eichinger} and refs. therein).

A second, and more ambitious goal of the project is to design non--con\-ven\-tio\-nal drugs which, instead of inhibiting the activity of the selected proteins, destabilize them by binding to the folding nucleus \cite{drug}, making them prone to proteolysis, that is, to degradation, usually by hydrolisis at one or more of its peptide bonds .  This approach is expected to be particularly interesting in the development of drugs against viral diseases, where it has been observed that essential proteins involved in the replication of the virus escape conventional drugs by inducing mutations in their active sites.  This road is not open to the virus to escape the action of non--conventional drugs since a mutation of an amino acid participating in the folding nucleus will lead to the denaturation of the protein.

The above simple considerations testify to the fact that a combined attack of the protein folding problem and of drug design through the interdisciplinary efforts of biologists, chemists and  physicists is likely to be a strategy which has a good chance of being succesful. The rewards of such a success in terms of basic research as well as of practical spin--offs are likely to be large.  In what follows we present the point of view of the physicist, placing special emphasis on minimal models and indicating how to extend the results to real proteins with the help of {\it first principle} all--atom quantal calculations.

\section{Main experimental facts}

The experiments which opened the way to the study of the protein folding problem have been those performed by Anfinsen \cite{anfinsen}. He studied the equilibrium conformation of small proteins like Ribonuclease A and Staphylococcal nuclease, changing cyclically the conditions of the solvent of the protein solution (pH, temperature, etc.). He observed that, independently on the past history of the solution, when this is driven back to physiological conditions, a protein (characterized by a given amino acid sequence) folds always to the same equilibrium conformation.  This result proves that the information about the unique equilibrium conformation of a protein\footnote{The uniqueness holds at the length scale of the overall structure of the protein. Experiments by Frauenfelder \protect\cite{frauenfelder} showed the existence of "conformational substates", that is fine--scale rearrangements of the native conformations. The energy scale at which these substates become relevant is $\sim 15$ meV (i.e. it is observable at $\sim 200$ K), and consequently we will not take them into account.} and about the pathway to reach it is completely encoded in the amino acid sequence.

Another feature which is common to most proteins is that their folding as well as the denaturation process is a highly cooperative process. The degree of folding of proteins (i.e., the relative population of the native state) in solution can be grossly assessed by circular dicroism experiments, which measure the rotation of the polarization of an incident laser beam induced by the protein. Since the native states of proteins are usually rich of motifs like $\alpha$--helices and $\beta$--sheets, and such motifs are able to rotate the polarization of an incident light beam, it is possible to measure the degree of formation of the motifs and approximate this with the degree of folding. Most proteins display a sudden transition in the relative population of the native state as the parameters of the system (e.g., temperature, pH of the solution, etc.) are varied from or towards the biological conditions. This result indicates that some cooperative mechanism, which involves all the parts of the protein, is acting to stabilize the native state. This behaviour is similar to the that of a physical system undergoing a first--order phase transition \cite{creighton2}.

The average folding time of proteins, that is the time needed to reach the native state starting from a random conformation, usually ranges from microseconds to seconds. Few exceptions span the order of magnitude of minutes, because of some lengthy, although not very common, structural change taking place inside one of the twenty types of amino acids (i.e., proline). The distribution of folding times is measured by stopped--flow, rapid quenching or flash photolysis methods \cite{fersht}, where one prepares the protein solution under denaturing conditions, reverts instantaneously the conditions of the solvent in order to start the folding process and then measures the degree of folding through optical techniques (e.g., circular dichroism). The resulting distribution of folding times is usually a single-- or multi--exponential function (in the sense that the concentration of folded proteins in the solution grows as a fuction of time with a single-- or multi--exponential behaviour). This indicates that a Poissonian processes, or a chain of (usually few) Poissonian processes, is at the basis of the folding mechanism.

As a rule, proteins are very tolerant to point mutations. Mutations in a large number of sites have little or no effect on the folding properties of the protein. For example, mutations in $61\%$ of the sites of Protein G causes an increase in the stabilization free energy of less than $20\%$ \cite{bakerG}. On the other hand, each protein displays some key sites which, if mutated, lead to a large destabilization of the native state.

\section{Evolutionary thermodynamics}

The number of proteins of known sequence and native conformation is of the order of some ten thousands. Nonetheless, the number of different topologies that the native conformations can assume is restricted to some hundreds. Since such sequences are the result of milions of years of evolution and in each step of evolution all the proteins must be good folders, under the risk of being selected out, the comparative analysis of sequences encoding for similar native structures can reveal information about their folding.

Naively speaking, if one compares all sequences folding to similar conformations and measures to which extent the type of amino acid is conserved in each site of the protein throughout all these sequences, it is possible to gain some insight into the role that the different sites play in the folding of that class of proteins. If a given site is always occupied by the same kind of amino acid, it is likely that that site plays an important role in the folding process.

\begin{figure}
\centerline{\psfig{file=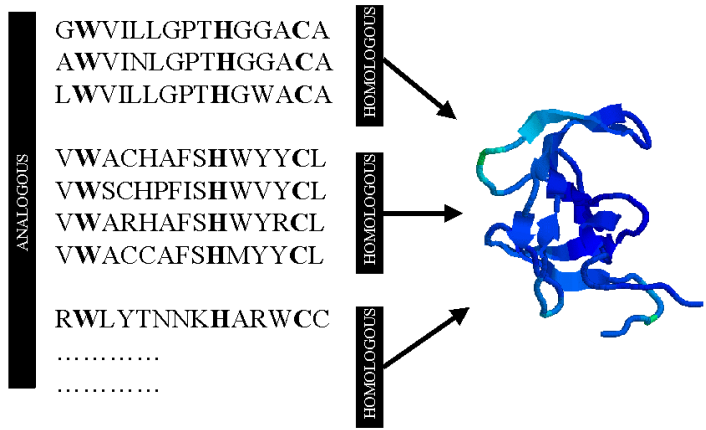,height=8cm}}
\caption{An example of aligment of sequences (left side) folding to the same native conformations (right side). Sequences displaying a high degree of similarity are defined as homologous and are grouped together in the figure. Pairs of sequences which display little similarity are defined as analogous (see text). Both analogous and homologous sequences share, as a rule, a small set of residues which are highly conserved (and which are indicated with bold characters in the left side of the figure).}
\label{sequences}
\end{figure}

In order to make this problem more quantitative, one can consider the space of sequences and study the subspace associated with a particular native conformation (cf. Fig. \ref{sequences}). A useful order parameter to study this space can be defined as the similarity between pairs of sequences
\begin{equation}
q_{seq}^{\alpha\beta}=\frac{1}{N}\sum_i^N \delta(\sigma_i^\alpha,\sigma_i^\beta),
\end{equation}
where the Kronecker delta is equal to 1 if $\sigma_i^\alpha=\sigma_i^\beta$ and zero otherwise.
Here $\alpha$ and $\beta$ label two sequences, while $\sigma_i$ indicates the type of amino acid occupying site $i$, while $N$ is the length of the sequence. The typical distribution $p(q_{seq})$ of the order parameter displays two peaks, one around $q_{seq}\approx 1$, corresponding to sequences which are different for few amino acids, and the other around $q_{seq}\approx 0.1$, corresponding to sequences whose similarity is comparable to that of random sequences \cite{rost}. The quantity defined above is similar to Parisi's replica order parameter for spin glasses, which is the paradigm of physical systems controlled by a disordered, complicated interaction (for details see ref. \cite{parisi}).

We will call "homologous" the pairs of sequences which belong to the large--$q_{seq}$ peak and "analogous" those belonging to the small--$q_{seq}$ peak. In order to study the conservation of protein sites, we will make use of analogous sequences only. The reason is that these better comply to the hypothesis of statistical independence. Since homologous sequences are clearly strongly correlated, the results of their statistical analysis suffer from errors due to the particular choice of the pairs of proteins (e.g., emoglobin of some animals belongs to a class of homologous proteins, which, for historical reasons, has been extensively studied). Consequently, these proteins are not statistically independent, and the conservation patterns arising from their study may result biased. Furthermore, the conservation pattern of homologs is likely to reflect the same functional requirements, like the ability to bind one given kind of molecules (e.g., to oxygen in the case of hemoglobin). While the conservation pattern of analogous families of proteins are likely to be also affected by functional requirements, the variety of binding requirements are expected to be less restrictive, thus leading to weaker correlations in the sequence ensembles.

For each class of analogous proteins it is possible to define the entropy per site, in the form
\begin{equation}
S(i)=-\sum_{\alpha=1}^{20}p_i(\alpha)\ln p_i(\alpha),
\end{equation}
where $p_i(\alpha)$ is the probability of finding an amino acid of kind $\alpha$ in site $i$. The probability $p_i(\alpha)$ is normalized in such a way that, for each site, $\sum_\alpha p_i(\alpha)=1$. This quantity can be simply calculated counting the different types of amino acids which occupy each site in the set of analogous sequences.  The site entropy indicates the degree of conservation of a site: the lower the entropy, the more conserved the site is. The largest value which the entropy $S(i)$ can assume is $-\ln (1/20)\approx 3$, corresponding to the case of uniform distribution of twenty kinds of amino acids.

\begin{figure}
\centerline{\psfig{file=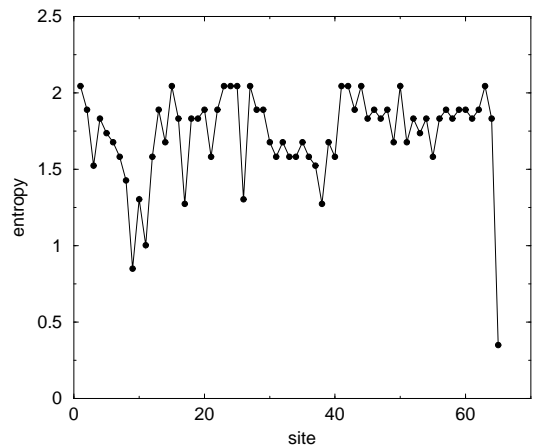,height=6cm,angle=-90}}
\caption{The entropy per site in the space of sequences, defined in Sect. 3, for the familiy of analogs of Chemotrypsin Inhibitor 2.}
\label{entseq}
\end{figure}

The analysis of the site entropy for a number of proteins \cite{mirny} shows that each class displays a small number of low--entropy (i.e., highly conserved) sites. An example is given in Fig. \ref{entseq}. The observation that the different sequences which build a class have different biological tasks and different active sites (i.e., the sites on the surface which are responsible for the enzymatic activity of the protein) rules out the possibility that conservation is solely associated with the biological function of the protein.  In ref. \cite{mirny} the authors conclude that the conserved sites are those which control the folding process of the proteins.

\section{Simplified models for protein folding}

A number of models to describe the folding process at various level of approximation have been proposed (cf. e.g. \cite{varenna} and refs. therein). In what follows we briefly review some of them. 

\subsection{Chemical--reaction model}

The simplest model used to describe the folding of proteins consists in summarizing all possible conformations in few macroscopic states, in the same way as it is done to describe chemical reactions (see e.g. ref. \cite{fersht}). Usually the thermodynamically relevant states are the unfolded state ("U"), which contains all the conformations where the protein chain is unstructured, the native state ("N"), which corresponds by definition to a single conformation\footnote{With the caveats discussed in Footnote 1.}, and possibly some intermediate states ("I$_1$", "I$_2$",...). The folding reactions is thus described by a chain of events like $U\rightarrow I_1\rightarrow I_2 \rightarrow ...\rightarrow N$ leading the protein from the unfolded to the native conformation (and, in some cases, being trapped aside in dead--ends). The different states are considered separated by free energy barriers. The definition of the model is then completed by assigning to each state and to the top of each barrier (usually called "transition state") a numeric value of the free energy.

The kinetics of a protein is then determined by Kramer's relation \cite{kramer}, which is a stochastic equation controlling the crossing of energy barriers in terms of the one dimensional energetic profile of the folding pathway \cite{hanggi}. For the simplest case of a two--state folding pathways ($U\rightarrow N$), the distribution of folding times which solve Kramer's Equation is a single exponential, whose characteristic time $\tau$ can be approximated by
\begin{equation}
\label{tau}
\tau=k\cdot \exp\left(\frac{\Delta F_{U-B}}{T}\right).
\end{equation}
The prefactor $k$ depends on the curvature of the energy profile and on the viscosity of the solution. The quantoty $\Delta F_{U-B}$ is the difference between the top of the free energy barrier ("B") and the free energy of the unfolded state, while $T$ is the temperature in energy units. If intermediates are present in the folding pathway, the distribution of folding time results in a sum of exponentials, each of them characterized by a characteristic time satisfying Eq. (\ref{tau}), where $ \Delta F_{U-B}$ is substituted by the free energy difference associated with the relative barrier.

Naturally, this model accounts for the cooperativity of the folding transition. It also describes the single-- and multi--exponential distribution of folding times (see Section 2). It can be used in connection with empirical free energy functions to predict the effect of mutations on the stability and kinetics of a given protein \cite{serrano1000}. On the other hand, it provides little insight into the molecular mechanism of the folding process. Moreover, the model describes the folding mechanism as a one--dimensional process, a simplification which is likely not to be very realistic.

\subsection{Entropy--based models}

The so called "new view" (cf. e.g. \cite{dill}) aims at understanding protein folding through a carefull description of the energy landscape associated with the conformational space.
In this class of models the protein chain is described at some degree of realism and one tries to characterize energetically, usually through computer simulations, a large number of conformational states along and around the folding pathway.

A simple model which follows this point of view is the Go model \cite{go}. In it, each amino acid of the protein can be described in terms of a full--atom representation at one extreme of detail, or as a structureless  spherical bead at the other extreme, depending on the computational cost one is prepared to afford. The potential function is a sum of two--body contact terms. Each term assumes the value $-1$ if the contact in question is present in the native conformation (which is assumed known) and zero otherwise. It reads
\begin{equation}
\label{gomodel}
U(\{r_i\})=-\sum_{ij}\Delta(r_i-r_j)\Delta(r_i^N-r_j^N),
\end{equation}
where $\{r_i\}$ denotes the cartesian coordinates of the atoms or of the amino acids, depending on the kind of description chosen, $\{r_i^N\}$ being the coordinates of the native conformation while $\Delta(r_i-r_j)$ is a contact function which assumes the value $1$ if the $i$th and $j$th atom (or amino acid) are closer than a contact threshold distance, $-\infty$ if they are closer than a hard--core threshold distance and zero otherwise. In other words, the potential function records how many native contacts there are in a given conformation, giving an energy $-1$ to each of them.

According to this model, the native state is, by definition, the ground state of the system and is also unique. Conformational samplings performed, for example, through a Monte Carlo algorithms \cite{metropolis} shows a cooperative folding transition of the kind observed experimentally (cf. Section 2). With some computational effort it has been possible to characterize the transition state, that is the state associated with the top of the main free energy barrier which separates the native from the unfolded state\footnote{In other words, the transition state describes the set of conformations for which the probability to fold or to unfold coincides and is equal to 1/2.}. For many small proteins, it has been found \cite{clementi,sh_crambin} that this state, for each protein, is characterized by a small number of well--defined contacts. Thus, the protein does not fold through a random collapse of the chain, but follows a well--defined sequence of steps.

The basic idea behind the Go model is that the geometry of the protein is the major determinant of its folding mechanism and, consequently, one can approximate the free energy associated with the conformational states through its entropic part. In this way one neglects the complexity of the interaction among the twenty kinds of amino acid and the possibility of building non--native interactions. Accordingly, the sequence of amino acid plays no role and there are no metastable states which can trap the chain.  
Notwithstanding all these caveats, the Go model has been important in emphasising the presence of a well--defined sequence of molecular events along the folding pathway.

\subsection{Energy--based models}

A different approach to the protein folding problem focus its attention on the energetic content of the free energy function. In particular, on the heterogeneity of the interaction arising from the presence of twenty kinds of different amino acids. It is known that physical systems displaying such an heterogeneity are associated, as a rule, with a rough energy landscape with many competing low--energy states \cite{parisi}. This is a picture incompatible with that of proteins, which must display a unique ground state, well separated from the others, and as few metastable states as possible. The purpose of these models is to understand what makes a protein, characterized by a well defined amino acid sequence, different from a generic heterogeneous system, whose paradigm is found in a random sequence of amino acids.

An important ingredient of this kind of models is the potential function. The simplest choice is that of a contact potential of the form
\begin{equation}
\label{hamilton}
U(\{r_i\},\{\sigma(i)\})=\sum_{ij}B_{\sigma(i)\sigma(j)}\Delta(r_i-r_j),
\end{equation}
where $r_i$ and $\sigma_i$ are the position and kind of the $i$th amino acid, $\Delta(r_i-r_j)$ is the contact function defined in connection with Eq. (\ref{gomodel}) and $B_{\sigma\tau}$ is the element if the 20$\times$20 interaction matrix which defines the iteraction energy between amino acids of kind $\sigma$ and $\tau$. A widely used interaction matrix has been determined by Miyazawa and Jernigan \cite{mj} from the statistical analysis of the contacts in a large database of known proteins, assuming that the frequency with which a given contact appears in the database measures the strength of the contact energy between the corresponding amino acids (cf. Table 1). This is done by calculating the probability $p_{\sigma\tau}$ of appearence of the contact between the amino acids of kind $\sigma$ and $\tau$, and assuming a Boltzmann--like relationship of the kind $B_{\sigma\tau}\sim -\log p_{\sigma\tau}$.

The starting point of energy--based models is the study of the thermodynamics of a random sequence, that is, of a chain of lineraly connected, random chosen monomers. The simplest model used to describe the thermodynamical properties of such sequences is the random energy model (REM)\cite{derrida}. In this model, the energy of each bond is assumed to be random, and not correlated with the energy of the other bonds. For a configuration with $\mathcal{N}$ contacts, the conformational energy is 
\begin{equation}
 E=\Sigma_{i}^\mathcal{N}\epsilon_{i},  
\end{equation}
where $\epsilon_i$ is the energy associated with the $i$th contact of the protein and is, by assuption, a value picked at random from the interaction matrix $B$.
Since we are dealing with random, uncorrelated values, the probability that a given configuration has energy E is given, if $N$ (number of beads) is large, by the central limit theorem\footnote{The central limit theorem states that if one has a large number of experiments which measure some stochastic variables (i.e., a quantity whose value is a number determined by the outcome of an experiment) then the probability distribution of the average of all the measurements approaches a Gaussian distribution. In other words, it   states that the probability distribution of the sum of independent stochastic variables approaches a Gaussian distribution as the number of variables increases.} 
\begin{equation}
P(E)=exp({\frac{-E^2}{2\mathcal{N}\sigma^2}}) 
\end{equation}
where $\sigma$ is the standard deviation of the interaction and where the distribution of contact energies is supposed to have zero mean.  The associated number of states is $n(E)=\gamma^{N}P(E)$ where $\gamma^{N} $ is the total number of conformations available to the chain with $\gamma$ ranging from 1.8 \cite{flory} to 2.2 \cite{orland} in the case of the simple cubic--lattice models discussed in Sect. 5, while it takes the value $4.8$ for more realistic models\footnote{The number of conformations {\it per monomer} $\gamma$ is equal to the effective number of nearest neighbours of a monomer. This number is lower than the actual number of nearest neighbours (which would be 6 for the cubic lattice, and $3\times 3=9$ for realistic models, where $3$ is the number of energy minima associated with each of the two torsional degrees of freedom of each amino acid. See \protect\cite{flory}) if one considers only the compact, thermodynamically relevant conformations.}\cite{flory}. The number $\mathcal{N}$ of contacts of the chain is, in average, equal to $N\gamma/2$ (the factor $1/2$ arising in order not to count twice the contacts), which, for the cubic--lattice model gives $N\approx \mathcal{N}$. Thus, one can write $n(E)$ as
\begin{equation}
\label{states_rem}
n(E)=exp{(-N(\frac{E^2}{2N^2\sigma^2}-ln\gamma))}.  
\end{equation}
It is seen that, if E$<E_{c}\equiv
-N\sigma(2\ln\gamma)^{1/2} $,
the exponent in Eq. (\ref{states_rem}) is negative. As a consequence, the number of states available with $E<E_{c}$ decreases exponentially with the length of the chain. For $E>E_{c}$ there are, in the limit of large $N$, many states, i.e,  
\begin{equation}
S(E)= 
N\ln\gamma-\frac{E^2}{2N\sigma^2}. 
\end{equation}
The quantity $E_{c}$ can thus be viewed as a threshold energy separating two regimes.
Eq. (\ref{states_rem}) expresses the fact that $E_{c}$ is the minimum conformational energy associated to a random heteropolymer.  Within the REM approximation, $E_{c}$ depends only on general features of the system (i.e. N and $\sigma$ and $\gamma$), and not on the details of the sequence. 

So far we have given a description of the REM based on the energy, that is a microcanonical description. It is often useful to give up to the precise control over energy and to describe the system in terms of the temperature 
\begin{equation}
\left.\frac{\partial S}{\partial E}\right|_{E=<E>}=\frac{1}{T},
\end{equation}
which sets the average energy $<E>$ but allows fluctuations about it, of the order $<E^2>-<E>^2=-\partial E/\partial (T^{-1})$ (i.e., a canonical ensemble description).  The existence of a lowest--energy--state ($E_c$) can be used to define a critical temperature
\begin{equation}
T_{c}=\frac{\sigma}{(2log\gamma)^{1/2}}.
\end{equation}
If one decreases the temperature below $T_c$, the system remains frozen in the last available state, i.e. $E_{c}$. In other words, the quantity $T_c$ sets the temperature scale of the system.

\begin{figure}
\centerline{\psfig{file=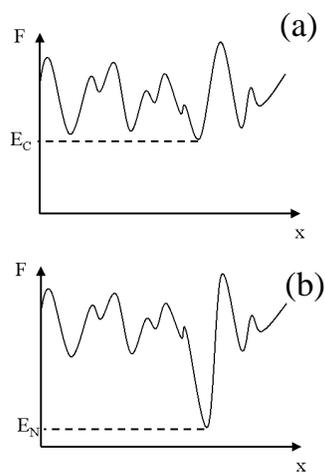,height=8cm}}
\caption{Sketch of the free energy landscape of (a) a random heteropolymer, (b) a selected protein. The quantity $F_c$ is the free energy of the lowest conformation of the random heteropolymer, while $F_N$ is that associated with the native conformation of the protein. The x--axis corresponds to a generic conformational coordinate. Note that, although this coordinate is one--dimensional for necessity of drawing, the conformational space of a protein is very--high dimensional.}
\label{landscape}
\end{figure}

Summing up, a random sequence displays a continuum of states, the lowest of which is $E_c$. While random sequences present a unique ground state this state is not well separated in energy from a large number of low energy states. Shakhnovich has shown (using a replica approach similar to that used for spin glasses \cite{parisi}) that these states belong to different energy valleys of a rough energy landscape, valleys correpsonding to very different conformations and separated by conspicuous energy barriers (cf. Fig. \ref{landscape}(a)) \cite{sh_analyt}. The resulting picture is that of an energy landscape characterized by a large number of competing low--energy states, and consequently displaying a thermodynamics very different from that of a protein (cf. Fig. \ref{landscape}(b)).

Also the kinetical features of a random sequence are quite different from those of a protein. The roughness of the energy landscape produces a myriad of metastable states which can trap the kinetics of the protein chain. In particular, it has been shown by Bryngelson and Wolynes studying the kinetics of the random sequences \cite{wolynes} that there is a temperature $T_g$ below which the kinetics is frozen, that is the protein hardly can escape from metastable states.  The temperature $T_g$ for the random energy models happens to be equal to $T_c$. As a consequence, in the range of temperatures where the low energy states are populated, the kinetics is frozen.

The question is then, how is it possible to find sequences displaying protein--like features? Necessary conditions for these sequences are 1) that they display a unique, zero--entropy ground state and 2) that the critical temperature $T_f$ below which the ground state is populated is somewhat higher than the temperature $T_g$ at which the kinetics is frozen, in such a way that the range of temperatures $T_g<T<T_f$ is suitable for folding.

It was shown by Shakhnovich \cite{sh_prl} that a sufficient condition to find good folders is to search for sequences whose native energy is well below $E_c$. Since the probability to find a random sequence with native energy below this threshold is exceedingly low,  sequences such that $E_N\ll E_c$ are likely to display a unique native state with large probability.
Moreover, such sequences will display a folding temperature $T_f$ higher than the kinetic freezing temperature $T_g$. 

\begin{figure}
\centerline{\psfig{file=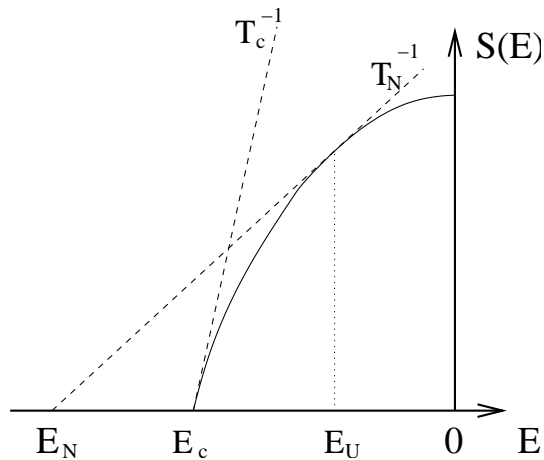,height=6cm}}
\caption{A sketch of the entropy as a function of energy for a selected sequence. The energy $E_N$ of the native state lies well below $E_c$. The critical temperature $T_c$ is the inverse slope of the tangent to the entropy in $E_c$. The folding temperature $T_f$ is tangent to the entropy at the energy $E_U$ corresponding to the unfolded state and passes through the native energy $E_N$ (the corresponding entropy being zero by definition).}
\label{gap}
\end{figure}

Let us first define $\delta=E_c-E_N$ as the energy gap between the native state and the lowest state belonging to the part of the spectrum described by the random energy, which is associated with all the conformations structurally different from the native one.  In ref. \cite{sh_spectrum} it was shown that the condition $T_f>T_g$ is equivalent to the presence of a positive gap $\delta$. According to the very definition of temperature ($\partial S/\partial E=1/T$), the critical temperature $T_g$ is the inverse slope of the tangent to the entropy evaluated in $E_c$ (see Fig. \ref{gap}). On the other hand, $T_f$ is defined by the condition $F_N=F_U$, where $F_N$ and $F_U$ are the free energies of the native and of the unfolded state (i.e., the state belonging to the part of the spectrum described by the random energy model and displaying a local minimum in free energy), respectively. Since the native state has, by definition, zero entropy, this condition becomes $E_N=E_U-T_fS_U$. $T_f$ is then the inverse slope of the straight line tangent to the parabolic line defining the entropy in the random energy model and having zero entropy in $E_N$. From Fig. \ref{gap} one can conclude that the condition  $T_f>T_g$ is equivalent to $\delta>0$.

Operatively, finding sequences with a large gap $\delta=E_c-E_N$ (compared to the energy scale $T_c$ of the system) can be done as follows \cite{sh_prl}: select a conformation to be the native, set the amount of the different kinds of amino acids, and minimizes the energy of the sequence by swapping the amino acids. In doing this, the energies of the unfolded conformations (and thus $E_c$) do not change, because they depend only on the amino acids composition, while the energy of the native conformation, which depends on the particular sequence, reaches values below $E_c$. This method can be implemented, for example, with a Metropolis Monte Carlo algorithm \cite{metropolis} in the space of sequences, performed at a low enough "selective" temperature $T_s$ (e.g., the temperature defined in the space of sequences, which has the physical meaning of evolutionary bias towards low energy sequences). Due to the size of the space of sequences, it is unlikely to be able to find the absolute energy minimum, but any sequence with $E<E_c$ will do the job.

To be noted that the design of good folders does not solve the protein folding problem, but the so--called inverse--folding problem, namely: given a target conformation, find the sequences which display this conformation as native state (i.e., non--generate, stable and, as we shall discuss in Sect. 5, kinetically accessible). Nonetheless, the systematic study of the folding of these designed sequences has opened the way to a {\it bona fide} solution of the protein folding problem, albeit still within the framework of minimal models of proteins (cf. Sect. 5). In any case, its extension to real proteins looks technically possible, provided that a reliable function to describe the potentials among amino acids is available.

\subsection{Molecular dynamics models}

An research tool whose importance has grown following the increase of computer power available is molecular dynamics simulation, where the protein is described in atomic detail and folding trajectories are generated making use of molecular dynamics algorithms. A possibility to carry out such a program consists in using an explicit description of the solvent molecules and integrate Newton's equations of motion. An alternative method, which allows to save some computational time, is to use Langevin's equations \cite{langevin}, taking into account the solvent in an implicit way.

There is a wide choice of realistic potential functions which can be used in connection with molecular dynamics simulations. Most of them, like GROMACS \cite{gromacs}, AMBER \cite{amber} and CHARMM \cite{charmm}, are obtained from chemical calculations of simpler molecules. The accuracy of these functions in describing the actual potentials among amino acids is controversial.

The fundamental drawback of this kind of calculations consists on the fact that, due to their complexity, it is only possible to simulate, within a reasonable amount of cpu time, few trajectories lasting for a tiny fraction of the overall folding time. For example, the most ambitious simulations performed to date consists in a single trajectory of 1 $\mu s$ of one of the smallest known proteins, namely villin \cite{duan}.

Although it is not possible to carry out a full folding trajectory, nor to collect meaningful statistics over different molecular events, one can still simulate the unfolding of protein chains, starting from the native conformation. The reason for this is that unfolding simulations can be performed at high temperature, thus decreasing the reaction times in an important way. Calculations of this type \cite{dinner} (using lattice models) have shown that the amino acids which are important for the unfolding mechanism are the same which control the folding.

In fact, in the study of the unfolding of, for example, Chymotrypsin Inhibitor 2 \cite{karplus}, which highlights that at the transition state (i.e. the state at the top of the free energy barier between the native and the unfolded states) only 25\% of the native contacts are formed, and two amino acids  control the kinetics, namely alanine--16 and leucine--49. This is in agreement with the results  obtained by studying the effects  point mutations have on the kinetics of the protein \cite{otzen}.

\section{Lattice models}

A powerful tool to study the physics of the folding mechanism of proteins is the lattice model. It is based on two approximations. First, the internal atomic structure of the amino acids is neglected and each of them is described as pointlike. Within this approximation, the entropy associated to the internal degrees of freedom of each amino acid is neglected and the force field created by each amino acid is regarded as isotropic.
The second approximation consists in locating the beads representing the amino acids on the vertices of a cubic lattice of unitary side length. Accordingly, the conformational degrees of freedom are discrete. This is very convenient from a computational point of view and makes conformational entropy easy to handle. Making use of this approximation, the small scale motion of the protein (i.e., the peptide bond vibrations) is neglected and the chain is constrained to have unrealistic angles between monomers ($\pi/2$, $\pi$ and $3\pi/2$). A more realistic choice is to use a fcc lattice (the average mean square of the difference between real proteins and their projection onto a fcc lattice is $\sim 1\AA$ \cite{park}), although calculations are slightly more complicated. Since the choice of the lattice does not change the underlying physics, in the following we will restrict to the use of a cubic lattice. The potential function used in these calculations is that introduced in Eq. (\ref{hamilton}).

The two ingredients which, notwithstanding the strong approximations, the model retains are the polymeric character of the protein (i.e., the fact that amino acids are linked into a chain) and the heterogeneity in the interaction, reflected by the interaction matrix $B_{\sigma\pi}$. These two ingredients are source of frustration, that is the impossibility for the system to satisfy all interactions at the same time \cite{toulouse}. Frustrated systems, as a rule, give rise to a rough energy landscape, of the kind described in Sect. 4.3 for random sequences. Sequences selected making use of the algorithm described also in Sect. 4.3 to fold to a unique native conformation are those which somewhat minimize the frustration of the system \cite{wolynes_frust}.

The study of the dynamics of these designed lattice--model proteins has uncovered a remarkably simple, strongly hierarchical picture of the folding process. Simulating the dynamics of the protein chain by means of , for example, a dynamical Monte Carlo algorithm (cf. Appendix), it is found that the whole process can be summarized as follows (cf. Fig. \ref{hierarchy})\cite{aggreg,jchemphys2}: 

\begin{enumerate}
\item Formation of local elementary structures (LES), built by amino acids close along the chain, very early in the folding process, i.e in times of the order of ${10^{-4}} $ the folding time (cf. Fig. \ref{hierarchy}(b)).  LES are, as a rule, stabilized by strong local contacts (dotted lines),

\item Docking of the LES in their native conformation (dashed lines, Fig. \ref{hierarchy}(c)) leading to the (post--critical, in the sense that it corresponds to a state beyond the top of the free energy barrier) folding nucleus,  that is, the minimal set of native contacts (summed of the contacts labeled by dashed and by dotted lines, Fig. \ref{hierarchy}(c)) which, once formed, guarantees fast folding. This event is interpreted as the overcoming of the main free energy barrier encountered by the protein in the folding process. In fact, after the formation of the folding nucleus the protein proceeds downhill, almost barrier-free toward the native state (cf. \cite{fersht}.

\item Relaxation of the remaining amino acids to form the corresponding native bonds which complete the folding process in times which are of the order of $10^{-3}$ the folding time (Fig. \ref{hierarchy}(d)).
\end{enumerate}

Such a hierarchy of events provides an effective mechanism according to which the entropy is squeezed out from the system (see Fig. \ref{squeeze}). In fact, once formed, the LES can be thought of as almost rigid structures. Consequently, the chain presents an effective length that is shorter than the actual one. Furthermore, the LES interact among themselves with energies which are much larger than those of single amino acids. Consequently, it is unlikely that LES form stable structures different from those for which they have been designed. In fact, such an event would imply the simultaneous optimization of a number of uncorrelated contacts, a highly unlikely scenario. As seen from Fig. \ref{squeeze}, the entropy is successively reduced  by the formation of the  folding nucleus.

\subsection{The impact of mutations}

\begin{figure}
\centerline{\psfig{file=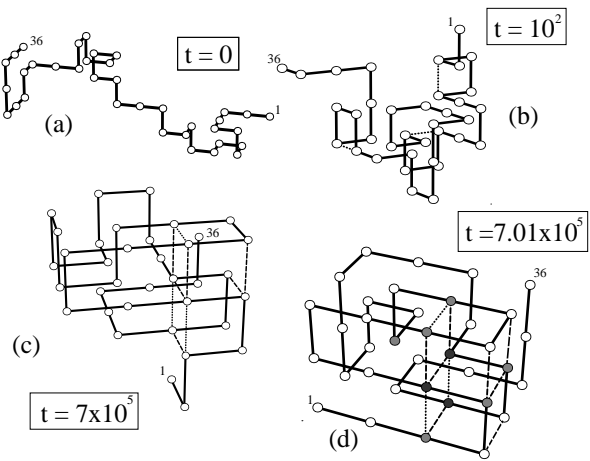,height=6cm,angle=-90}}
\caption{The hierarchy of folding events for a model protein consisting of 36 amino acids, whose native state enegy is $-17.13$ in the appropriate units ($RT_{room}=0.6\frac{cal}{mol}$), to be compared with $E_c=-14$. From a random coil conformation (a), local elementary structures are formed very early in the folding process, that is, after about $10^{2}$ Monte Carlo steps (b); next the folding nucleus forms (c) and the remaining native contacts are formed to complete the folding process (d). Black, grey and white beads indicate amino acids occupying hot, warm and cold sites (cf. Sect. 5.1).}
\label{hierarchy}
\end{figure}

\begin{figure}
\centerline{\psfig{file=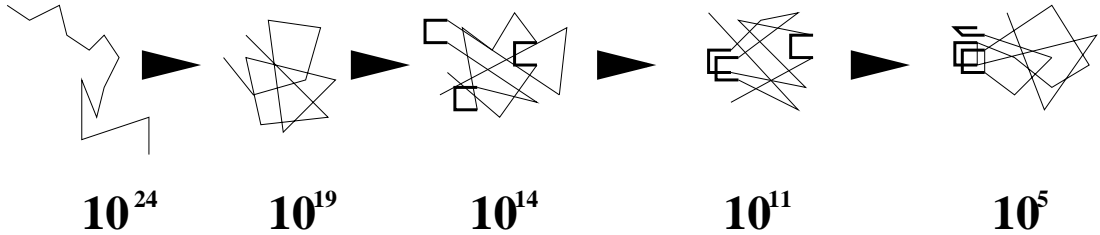,height=3cm}}
\caption{Steps by which sequence shown in Fig. \protect\ref{hierarchy} reduces its entropy.  For a random coil there are $10^{24}$ possible conformations. As the  system collapses to a random globule there are $10^{19}$ conformations available. Successively, LES are formed ($10^{14}$ conformations). The formation of the contacts between two of the three LES reduces the number of available conformations to $10^{11}$ while when the nucleus is completely formed the system has to search among  $10^{5}$ conformations. }
\label{squeeze}
\end{figure}

An important feature which makes protein-like sequences different from random heteropolymers is their behaviour with respect to mutations. In fact, random heteropolymers are very sensitive to mutations, in the sense that a mutation usually leads to a very different target conformation in the compaction process.  For protein-like sequences this is, as a rule, not the case. In fact a protein can undergo many mutations without changing its  native state conformation.  On the other hand, mutations made in special sites usually lead to a complete misfolding or to a great destabilization of the native state. These thoretical results agree with those found in studies of protein engineering in real proteins \cite{fersht}.
To make this point clearer we report the result of a study of the impact of point mutations on the designed sequence composed of 36 monomers whose folding mechanism is displayed in Fig. \ref{hierarchy} \cite{jchemphys1}.  From this study it was found that mutated sequences can be classified into three groups:

\begin{enumerate}
\item sequences that still fold to the native state,

\item sequences which fold to a unique compact structure, usually similar but not identical to the native one,

\item sequences which do not fold to a unique conformation.
\end{enumerate}

An analysis of the resulting sequences reveals that the impact of a mutation is dependent on the local change in energy induced on the native state, that is, 
\begin{equation}
 \Delta{E}_{loc}[m(i)\rightarrow m^{'}(i)]=\Sigma_j(U_{m^{'}(i)m{(j)}}-U_{m(i)m{(j)}})\Delta(|r_i-r_j|),
\end{equation}
where the amino acid $m$ at position i is substituted by the amino acid $m^{'}$. If $\Delta{E}_{loc}$ is small (compared to $m$) the mutation has no effect on the thermodynamics of the protein, but if large, it denaturates the protein.  According to the value of  $\Delta{E}_{loc}$ averaged over all nineteen possible mutations, the different sites of a native conformation can be classified as (cf. Fig. \ref{hierarchy}(d)):

\begin{enumerate}
\item hot: sites very sensitive to point mutations (black beads in Fig \ref{hierarchy}(d)), characterized by a high value of  $\Delta{E}_{loc}$ as compared to ($\sigma$=0.3). If a hot site undergoes a mutation, the resulting sequence becomes as a rule denaturated,

\item  warm: the resulting sequence either folds to the same structure or to a structure which is quite similar to it, occupying the native state with a  reduced probability as compared to the original sequence (grey beads in Fig. \ref{hierarchy}(d)).

\item cold: sites which are not sensitive to point mutation (white beads in Fig. \ref{hierarchy}(d)).
\end{enumerate}

LES are stabilized by hot and warm sites. This is the reason why mutation of type (1) or (2) can affect in an important way the folding ability of the sequence.

\subsection{A solution to the protein folding problem}

With the help of the results discussed above, a strategy known as the three step strategy (3SS) was developed, which allows one to predict the three--dimensional native conformation of a model protein from its amino acid sequence \cite{1d3d}, provided the contact energies acting among the amino acids, and which was used to design the protein, are known.   The algorithm consists of three steps, namely 1) finding good candidates for the role of local elementary structures, 2) finding good candidates for the folding nucleus and 3) finding the native conformation relaxing  the residues not participating in the folding nucleus. This algorithm is based on the hierarchical sequence of events that allows the chain to fold fast and works because at each step only a limited portion of the configurational space has to be searched through (cf. Fig. \ref{hierarchy}).

In what follows we briefly discuss the 1D$\rightarrow$3D algorithm and apply it to a representative example of notional proteins.  \begin{itemize} \item {\bf Step 1:}  finding  of LES which govern the folding process. Elementary structures are called  "closed" or "open", depending whether they contain interactions inside themselve (outside from the peptidic bond), or not. In fact, in some cases short fragments of the chain, stabilized only by the peptide bond, play the role of LES.  Examples of closed LES are provided by LES 3-6, LES 11-14 and LES 27-30 (cf. Fig. \ref{hierarchy}(b)), structures stabilized by the native contacts drawn in terms of dotted lines.  In keeping with this classification of LES, the present step  is composed of two substeps,

{\bf Substep 1a:} Finding the open elementary structures.  For each substring of the sequence, starting at monomer $i$ and ending at monomer $j$ ($0<i<j<N$), we define the density of energy
\begin{equation}
\epsilon_s=\frac{1}{j-i}\sum_{i\leq l\leq j}
\min_{k \in\hspace{-0.15cm}|\; (i,j)} U_{m(l)m(k)},
\end{equation}
where $U$ is the matrix of contact energies used to design the notional protein. In other words, $\epsilon_s$ is the average energy with which each element of the substring $(i,j)$ interact with the rest of the chain. The substrings which are good candidates to be open elementary structures in the  folding process have low values of $\epsilon_s$. Among such   substrings we select those with values of $\epsilon_s$ lower than a threshold $\epsilon_s^*$.

{\bf Substep 1b:} Finding the closed elementary substructures.  For this purpose we evaluate, for each pair of monomers $i$ and $j$,  the function
\begin{equation}
p(i,j)=\frac{\exp(-U_{m(i)m(j)}/T_{eff})}{(j-i)^\rho},
\end{equation}
where $T_{eff}$ is an effective temperature which we set equal to the standard deviation of the interaction matrix $U$ (e.g. $\sigma=0.3$ for the case of the contact matrix displayed in Table 1).  This function has been chosen in order to maximize the attraction between amino acids and minimize their distance.  The exponential factor $\rho=1.7$ reflects the ratio between the number of conformations associated with the formation of a contact and the total number of conformations.  If a substructure contains more than one interaction, the values of $p$ associated with the different interactions are to be multiplied together.  As possible (closed) local elementary structures, we select those composed of mononomers $i,i+1,...,j-1,j$ and with $p(i,j)>p^*$, where $p^*$ is a threshold value. 
\item {\bf Step 2:} Finding the folding nucleus. All the elementary structures (let $S$ be the total number of such structures) found in steps 1a and 1b are moved in space and the conformational spectrum is found. This is done selecting all possible choices of $1, 2,..., S$ local elementary structures, giving them all possible relative conformations and making a complete enumeration of their reciprocal positions in space.  The  conformations with lower energies are selected as possible candidates for the (post--critical) folding nucleus of the protein.

\item {\bf Step 3:} Relaxing the remaining monomers around the folding core.  This can be done through a complete enumeration of all the conformations displaying a given nucleus as their number is   rather low ($\sim 10^4$ for a 36mer). Another way, which we found computationally attractive is to use a low--temperature Monte Carlo relaxation simulations, keeping fixed the monomers belonging to the folding nucleus. \footnote{In some cases the system is non ergodic, in the sense that from a given starting configuration it is not possible to reach all other configurations (with the folding core formed and fixed). In such  cases several relaxation simulations  are performed starting from different conformations (with the folding core formed and fixed).In keeping with this fact, the folding nucleus of a notional protein could be required not to be exceedingly stable, so as to avoid long--lived metastable states {\it en route} to folding.} The (single) totally relaxed conformation with energy lower than $E_c$ is the native conformation of the protein.
\end{itemize}

\begin{figure}
\centerline{\psfig{file=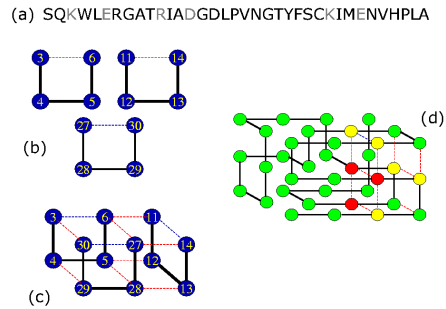,height=5cm,angle=-90}}
\caption{The primary structure of S$_{36}$ (a), its LES (b), the folding nucleus(c) and the native state conformation (d). To be noted that the amino acids composing the LES 3-6, 11-14 and 27-30 are KWLE, RIAD and KIME, respectively. Making use of the MJ contact matrix (cf. Table C1 App.C) $U_{KE}$=-0.97, $U_{MW}$=-0.60, $U_{LI}$=-0.41, $U_{IA}$=-0.22, $U_{ER}$=-0.74 and $U_{KD}$=-0.76 one obtaines that the interaction energies between the LES (3-6)-(27-30), (3-6)-(11-14) and (27-30)-(11-14) are -1.92, -1.15 and -0.98 respectively.}
\label{1d3d}
\end{figure}

\begin{figure}
\centerline{\psfig{file=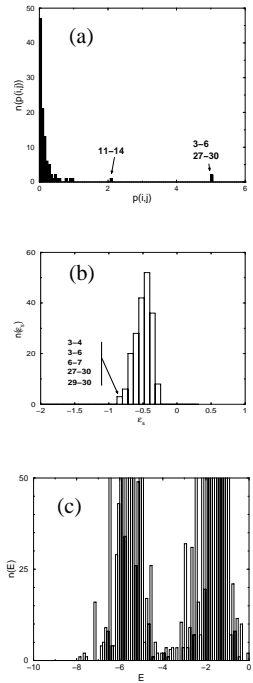,height=9cm}}
\caption{{\bf (a)} The distribution of the parameter p(i,j) (cf. Eq. (14)),whose maximization allows to find the closed elementary structures. {\bf (b)} the distribution of the energy density $\protect\epsilon_s$ (cf. Eq. (13)), employed to find open elementary structures.  {\bf (c)} The distribution of the energies associated with the possible folding nuclei of sequence S'$_{36}$,  build of the elementary structures 3--4--5--6, 11--12--13--14 and 27--28--29--30.}
\label{1d3d_b}
\end{figure}

The algorithm described above was tested with success on representative examples of lattice designed proteins. Below we discuss results concerning the designed sequence S$_{36}$ (cf Fig. \ref{1d3d}(a)), which folds into the native structure shown in Figs. \ref{hierarchy}(d) and \ref{1d3d}(d), displaying a native energy $E_n=-16.5$, much lower than the threshold energy $E_c=-14$ (thus $\delta/\sigma=(E_c-E_N)/\sigma\approx 8$).  In Fig. \ref{1d3d_b}(a) we display the distribution of values of $p(i,j)$.  Three bonds have a p--value which is remarkably larger than that associated with the rest of the possible bonds of the protein, and consequently are good candidates for stabilizing closed local elementary structures. The distribution of values of $\epsilon_s$, displayed in Fig. \ref{1d3d_b}(b), shows a single peak, whose lowest points are associated with the same sites already involved in the closed elementary structures. It is thus likely that open elementary structures do not play any noticeable role in the folding process of S$_{36}$.  We thus search for a folding nucleus composed of the LES $(3,4,5,6)$, $(11,12,13,14)$ and $(27,28,29,30)$, stabilized by the contacts $3-6$, $11-14$ and $27-30$.  A complete enumeration of all the conformations built out of these three elementary substructures gives the energy distribution displayed in Fig. \ref{1d3d_b}(c). The most stable of these conformation has energy $-7.81$ and is, in fact, the actual folding core (cf. Fig. \ref{1d3d}(c)). The relaxation of the other amino acids around it gives the right native conformation, with energy $E_n=-16.50$.  The next low--energy conformations built of the three elementary substructures have energy $-7.75$, $-7.68$ and $-7.68$. The relaxation of the other residues around these temptative folding nuclei lead to "native" energies energies $-12.40$, $-12.58$ and $-14.05$, respectively. The first two of them are larger than $E_c=-14$, so they correspond to states which belong to the set of structurally dissimilar conformations ($q<<1$) to the native conformation we are searching.  The last of them has an energy just below $E_c$. Although it can hardly be confused with the native conformation, it corresponds to a metastable state which can slow down the folding process.

\section{Design of non conventional drugs}

Drugs perform their activity by either activating or inhibiting some target component of the cell. In particular, many inhibitory drugs bind to an enzyme and deplete its function by preventing the binding of the substrate. This is done by either capping the active site of the enzyme   (competitive inhibition) or by binding to some other part of the enzyme  to the end of leading to a structural change which makes the enzyme unfit to bind the substrate (allosteric inhibition).  The two main features that inhibitory drugs must have are efficiency and specificity. In fact, it is not sufficient that the drug binds to the substrate and reduces efficiently its activity; it is also important that it does not interfere with other cellular processes, binding only   to the protein it was designed for. These features are usually accomplished by designing drugs which mimic the molecular properties of the natural substrate. In fact, the pair enzyme/substrate have undergone milions of years of evolution in order to display the required features, and consequently the more similar the drug is to the substrate, the lower the probability that it interferes with other cellular processes. 
 
Something that this kind of inhibitory drugs usually cannot do is to avoid the development of resistance, a phenomenon which is typically related to viral protein targets, in particular those of retroviruses which, carrying their genetic information on a single RNA strand, display a replication mechanism prone to errors (mutations). Under the selective pressure of the drug, the  target is often able to mutate the amino acids at the active site in such a way that the activity of the enzyme is essentially retained, while the drug is no longer able to bind to it. An important example of drug resistance is connected with that displayed by patients affected by the HIV virus (AIDS). In this case one of the target proteins, the HIV--protease, is found to mutate its active site so as to elude the effects of conventional drugs within a period of 6--8 months after the starting of the therapy.

Making use of the insight obtained from the study of model protein folding and of the 3SS discussed above, the possibility emerges of designing drugs which interfere with the folding mechanism of the target protein, destabilizing it and making it prone to proteolisis\footnote{That is, the cleavage into the original aminoacids of misfolded proteins. Proteolisis is operated by a number of enzymes which are quite ubiquitous in cells and whose function is to "clean up" the cell from non--functional proteins.}. Furthermore exploiting the hierarchical folding mechanism of proteins, it is possible to design drugs which not only are efficient, and specific but which, at the same time do not suffer from the upraise of resistance.

As shown above, local elementary structures are the building blocks that make up the folding nucleus. To perform their job in an effective way  LES interact strongly only with their complementary structures, avoiding the formation of metastable states. This fact suggests that peptides with the same sequence as LES could be used at profit as  drugs.  These peptides would interact strongly only with the LES of the protein, and consequently block their assembling.  To assess the correctness of these statements, the effect of these peptides (which will be shortened as p--LES), on the folding ability of a  variety of lattice designed proteins of different length have been studied. 

The central quantity in this study is the parameter $q$ which measures the similarity between a configuration at time t and the native state.  To each configuration $\Gamma_\alpha$ is associated a contact map $\Delta_{i,j}=\Delta(\mid{r_i-r_j}|)$.  Given a map $\Delta_{i,j}(\Gamma_\alpha)$ and the map relative to the native state $\Delta_{i,j}(\Gamma_N)$, the similarity parameter $q$ is defined as, 
\begin{equation}
q(\Delta_{i,j})=\frac{\Sigma_{i<j}\Delta_{ij}(\Gamma_{\alpha})\Delta_{ij}(\Gamma_{N})}{\Sigma_{i<j}\Delta_{ij}(\Gamma_{N})}.
\end{equation}
A typical simulation involves a designed protein sequence  and a number of shorter chains, whose length lies within the range of 2-12 residues. The peptides used during the simulation are indicated as p--LES n'-m', where n' and m' are the first and the last monomer of the p-LES, following the numeration of the protein sequence.  Their activity has been checked against that of non--specific  short peptides (denoted p n'--m') built of pieces of the protein not belonging to LES, whose first and last amino acid are n' and m', respectively.
  
At each Monte Carlo step of the simulation of the folding of the designed protein in the presence of $n_p$ peptides, one of the $n_p+1$ chains (i.e., either the designed protein or one of the peptides) is picked up with equal probability. Then a site of the selected chain is chosen with a probability $1/L$, where $L$ is the length of the chain. A move of the type displayed in Fig. \ref{mc_moves} (cf. Appendix) is then attempted. We let the system evolve for about $2\times 10^8$ Monte Carlo steps, recording the value of the similarity parameter $q$ at every 1500 steps. Making use of these values the normalized probability function $p(q)$ is constructed. The population of the native conformation is defined as the fraction of the probablity function with $q>0.7$, that is $\int_{0.7}^{1}p(q)dq$. The value $0.7$ has been chosen as it corresponds to the minimum in $p(q)$ separating the peaks associated with the unfolded and with the folded phases of the isolated protein (cf. black continuous curve in Fig. \ref{drug3-6}).

\subsection{Destabilizing effects of p--LES}

We shall first discuss the p-LES strategy on the test sequence $S_{36}$.  The simulations have been performed at the folding temperature, at which the population of the native state is $\frac{1}{2}$. Although this is quite a high temperature from the biological point of view, it allows to compare results obtained studying different lattice designed proteins, displaying different thermodynamical properties on equal footing.  In the case of S$_{36}$ the folding temperature is, in the units we are using ($RT_{room}=0.6\frac{Kcal}{mol}$), $T=0.24$.

\begin{figure}
\centerline{\psfig{file=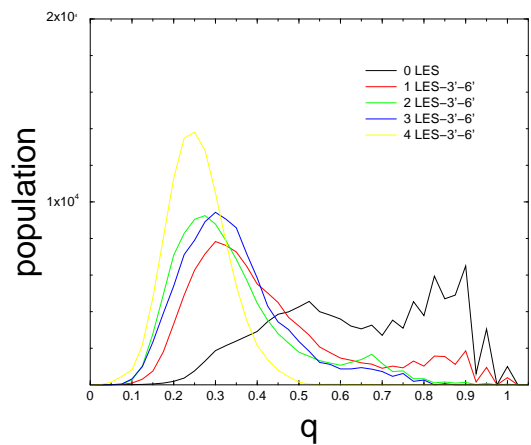,height=6cm,angle=-90}}
\caption{Equilibrium population of S$_{36}$ folding in the presence of $n_{p}$
\label{drug3-6}
p-LES 3'-6'.}
\end{figure}

\begin{figure}
\centerline{\psfig{file=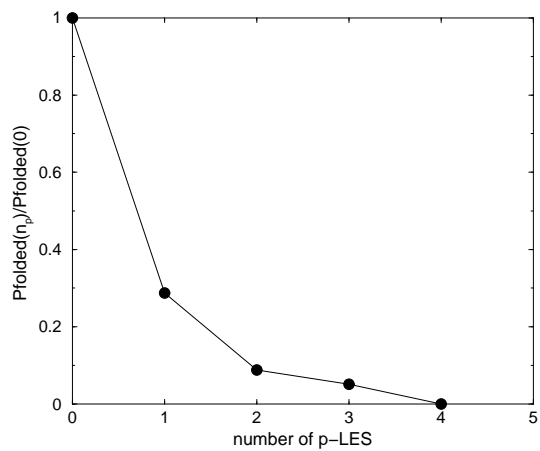,height=6cm,angle=-90}}
\caption{The relative occupancy $\frac{P_{folded}(n_p)}{P_{folded}(0)}$ of the native state shown in Fig. \protect\ref{1d3d}(d) by the sequence S$_{36}$ as a function of the number $n_{p}$ p-LES 3'--6'.}
\label{drug3-6b}
\end{figure}

We now study how the presence of a number $n_{p}$ of peptides of different types  which correspond both to p-LES and to non p-LES (as a check), affect the folding of the designed protein S$_{36}$.   As mentioned above, the folding and stability of the sequence $S_{36}$ is controlled by three LES, namely LES 3-6, LES 11-14 and LES 27-30.  The equilibrium distribution of q for the protein S$_{36}$ in presence of a number $n_{p}$ of p-LES of kind 3-6 is displayed in Fig. \ref{drug3-6}.  The degree by which the protein is hindered of reaching the native state in the presence of the peptide  3'-6' is shown in Fig. \ref{drug3-6b}, which displays a monotonic decrease of the population of the native state, reaching essentially zero for $n_{p}=4$.
   
\begin{figure}
\centerline{\psfig{file=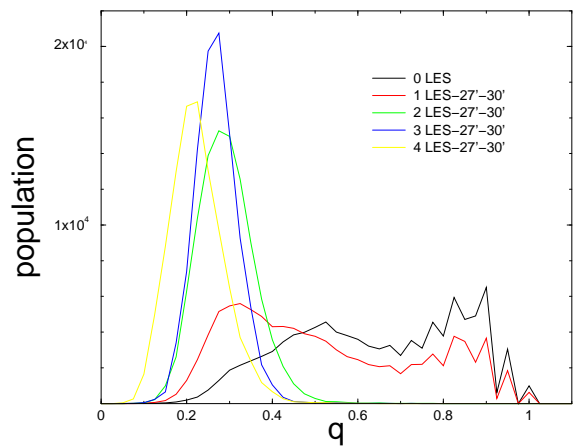,height=6cm,angle=-90}}
\caption{Equilibrium population of S$_{36}$ in presence of $n_{p}$ 27'--30' peptides.}
\label{drug27-30}
\end{figure}

\begin{figure}
\centerline{\psfig{file=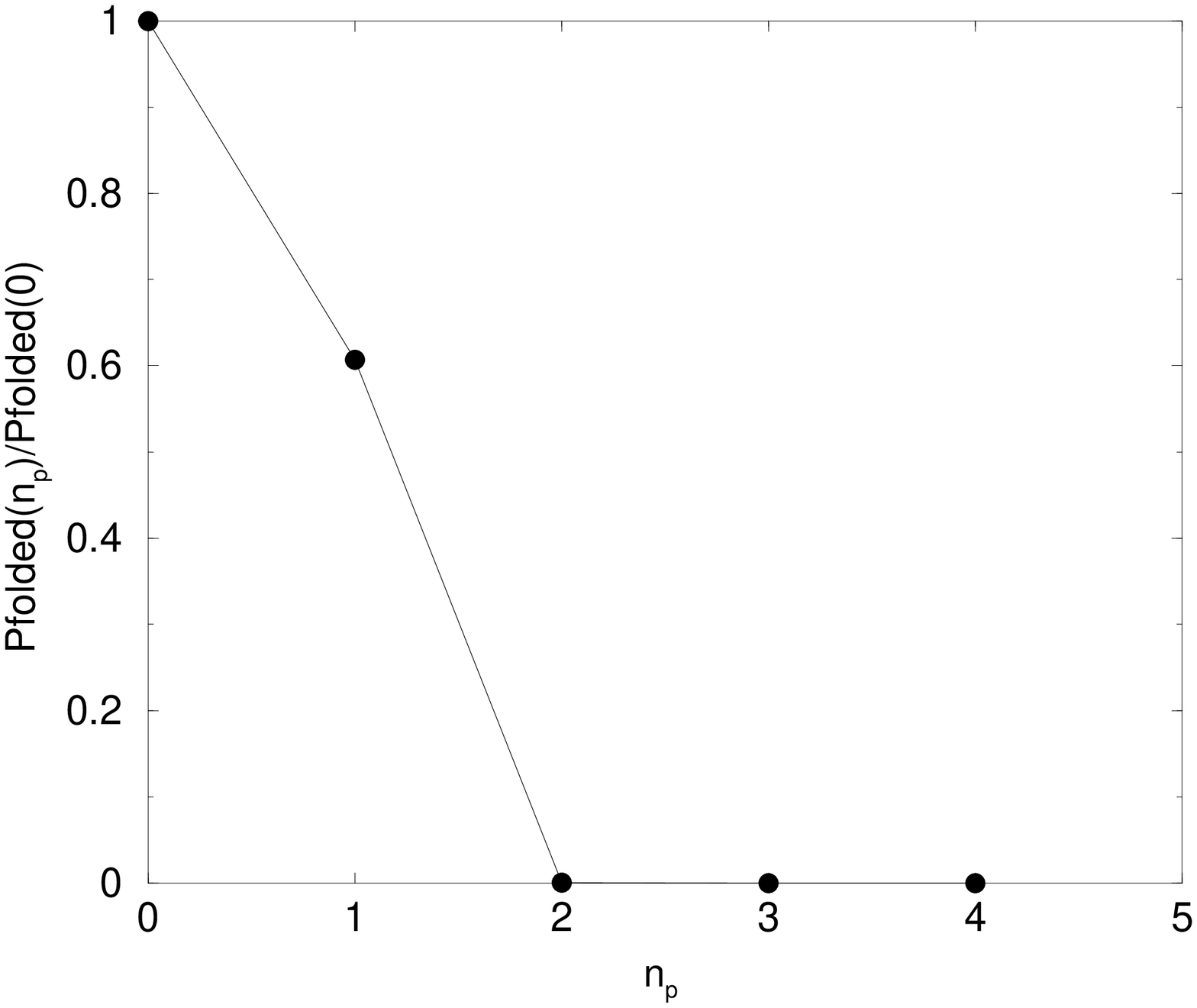,height=6cm,angle=-90}}
\caption{The relative occupancy $\frac{P_{folded}(n_p)}{P_{folded}(0)}$ of the native state shown in Fig. \protect\ref{drug27-30}(d) by the sequence S$_{36}$ interacting with $n_{p}$ p--LES 27'--30'.}
\label{drug27-30b}
\end{figure}

\begin{figure}
\centerline{\psfig{file=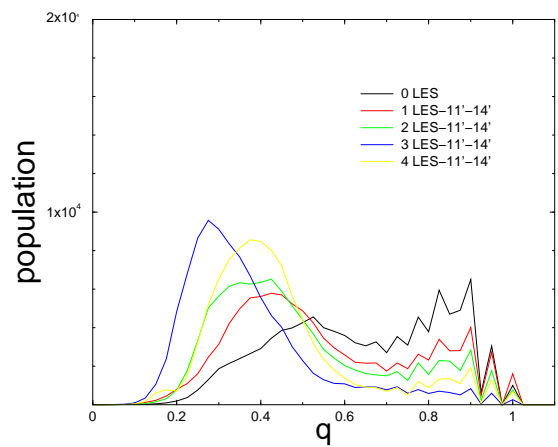,height=6cm,angle=-90}}
\caption{Equilibrium population of S$_{36}$ in presence of $n_{p}$ 11'--14' peptides.}
\label{drug11-14}
\end{figure}

\begin{figure}
\centerline{\psfig{file=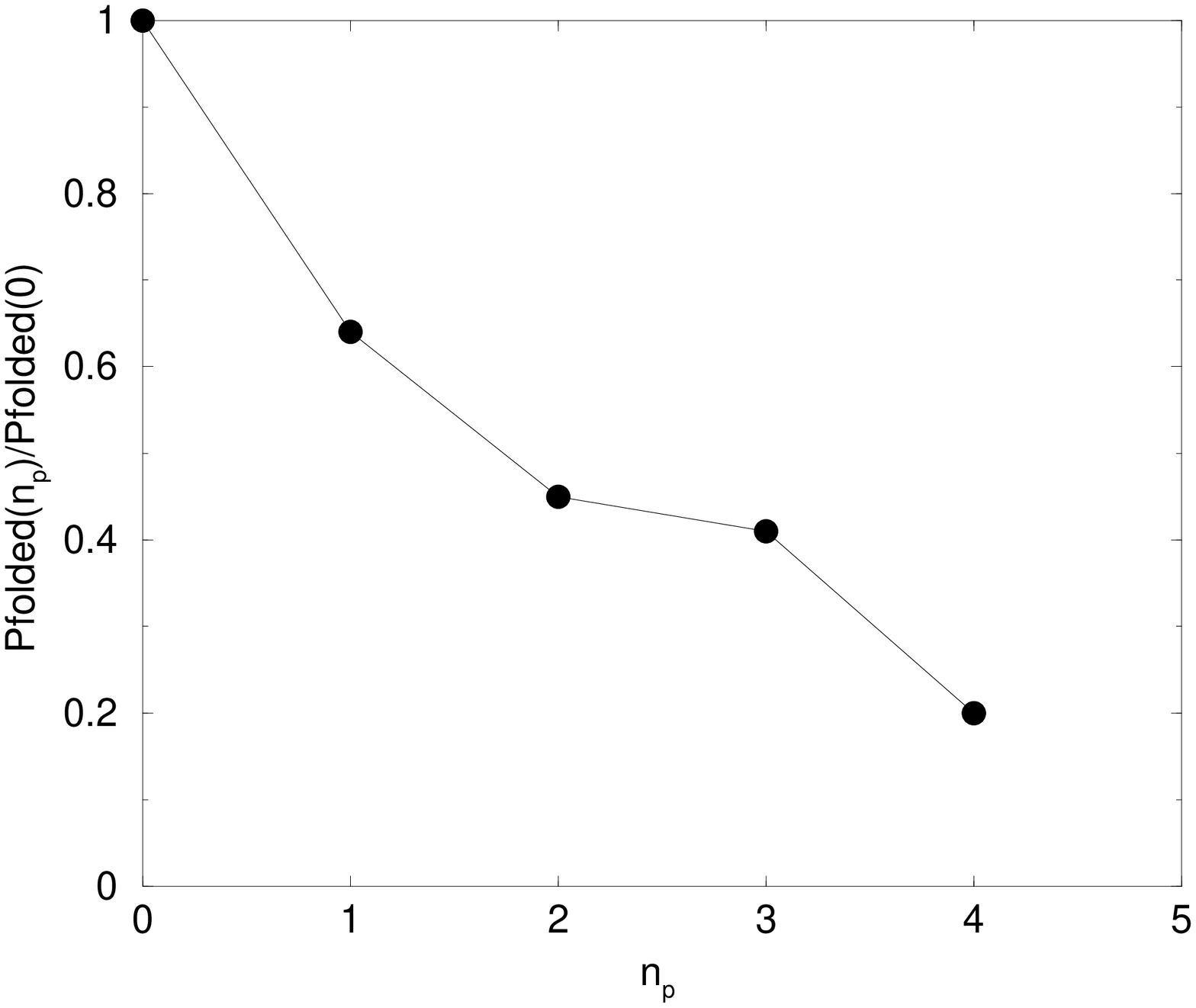,height=6cm,angle=-90}}
\caption{The relative occupancy $\frac{P_{folded}(n_p)}{P_{folded}(0)}$ of the native state shown in Fig. \protect\ref{drug11-14}(d) by the sequence S$_{36}$ interacting with $n_{p}$ p--LES 11'--14'.}
\label{drug11-14b}
\end{figure}

Similar to the previous case, the equilibrium distribution of states of S$_{36}$ in presence of $n_{p}$ p--LES 27'--30' and 11'-14' are displayed in Figs. \ref{drug27-30}--\ref{drug27-30b} and \ref{drug11-14}--\ref{drug11-14b}, respectively.  The effect of p-LES 27'-30' is stronger than that of the p--LES 3'-6', totally destabilizing the protein (i.e., the value of $\int_{0.7}^{1}p(q)dq$ is essentially zero) already at $n_{p}=2$.  On the other hand p--LES 11'--14' appears to be  somewhat less effective than the other two p-LES in blocking the folding of S$_{36}$ (cf. Fig. \ref{drug11-14b}).  The results shown above essentially do not change if instead of starting the folding simulations of S$_{36}$ with a number $n_{p}$ of p--LES with S$_{36}$ in a denatured (elongated) conformation, the simulations are started with S$_{36}$ in the native conformation, as was expected from the fact that we are studying the equilibrium properties of the system, and that proteins display important fluctuations around the native conformation. 
From what has been shown so far it is clear that the presence of one or more p-LES leads to an important destabilization of S$_{36}$.  A question which now arises is : what would happen if instead of peptides of type p-LES one uses peptides (eventually built out again of four residues) which corresponds to segments of the designed sequence not belonging to the folding nucleus ? To answer this question, simulations have been carried out in which the folding of S$_{36}$ is studied in the presence of a number $n_{p}$ of peptides p 8'--11' or p 30'--33'.

\begin{figure}
\centerline{\psfig{file=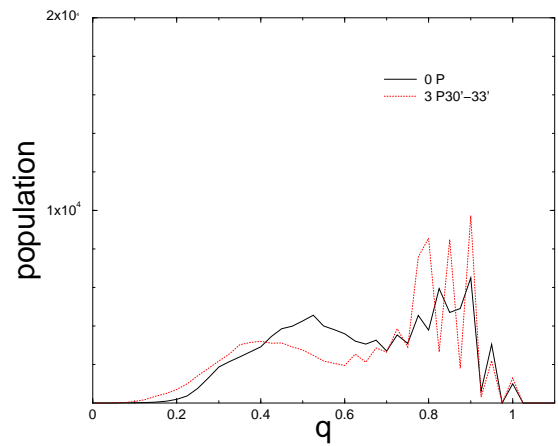,height=6cm,angle=-90}}
\caption{Equilibrium population of S$_{36}$ folding in presence of $n_{p}$ p 30'--33'.}
\label{drug30-33}
\end{figure}

\begin{figure}
\centerline{\psfig{file=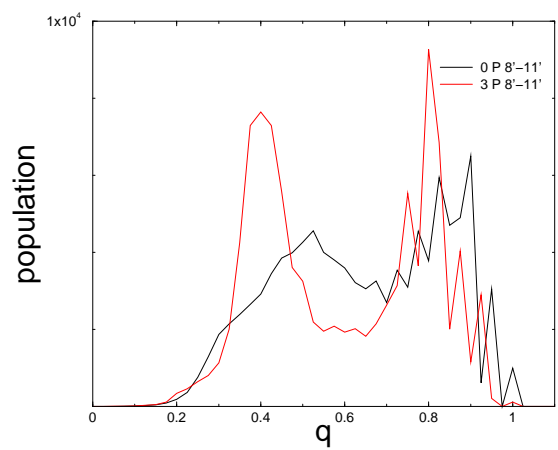,height=6cm,angle=-90}}
\caption{Stability of the sequence S$_{36}$ in presence of $n_p$ peptides which interact with the part of the protein with which the amino acids 8-11 of S$_{36}$ interact in the native state.  The contact energies of p 8'--11'have been increased so that this interaction is as strong as the average interaction energy of LES among themselves.}
\label{drug8-11mod}
\end{figure}

A single peptide of type 8'--11' seems to slightly increase the stability of S$_{36}$ while three destabilize it by a very small extent, both effects being only marginal. A similar result is found for p 30'--33' (cf. Fig. \ref{drug30-33}) To check whether these results are due to the fact that the interaction of peptides p 8'--11' and  30'-33' with the protein is much weaker than that associated with p-LES, we have increased the interaction of 8'--11' and 30'--33' with S$_{36}$, so as to mimic the interaction energies typical of p-LES.  In particular, in the case of p 8'--11' the interaction of the contacts 8'--21, 9'--22' 10'--15 and 11'--14' were increased to the average value of the contact energy acting between two LES.  The results are similar to those obtained using the original peptide 8'--11', indicating again the lack of destabilizing effects (cf. Fig. \ref{drug8-11mod}).

\begin{figure}
\centerline{\psfig{file=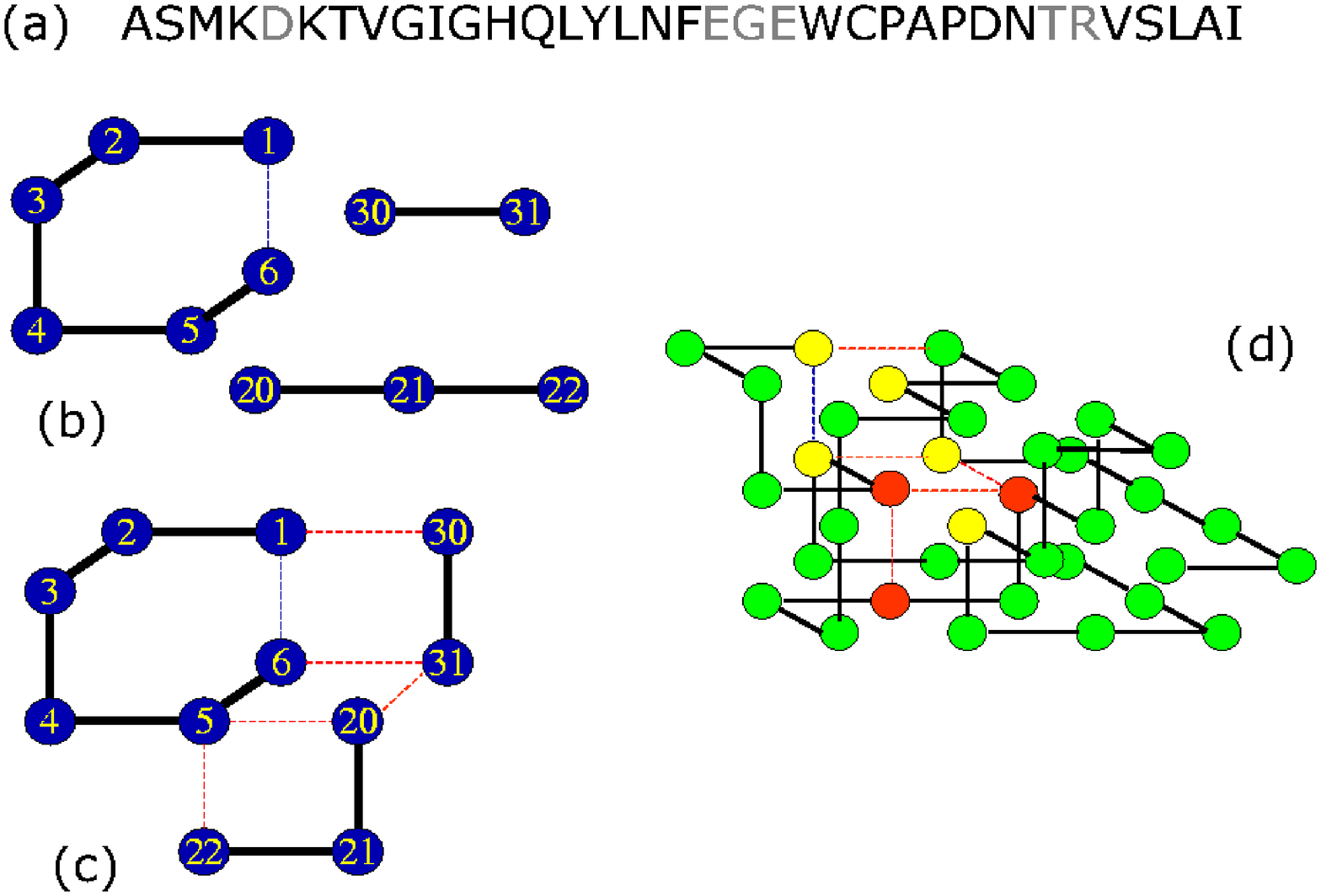,height=6cm,angle=-90}}
\caption{(a) Primary structure of the designed 36-mer \protect\cite{sh_prion}, its LES (b), the folding nucleus (c) and the native state conformation (d).}
\label{nonlocal}
\end{figure}

\begin{figure}
\centerline{\psfig{file=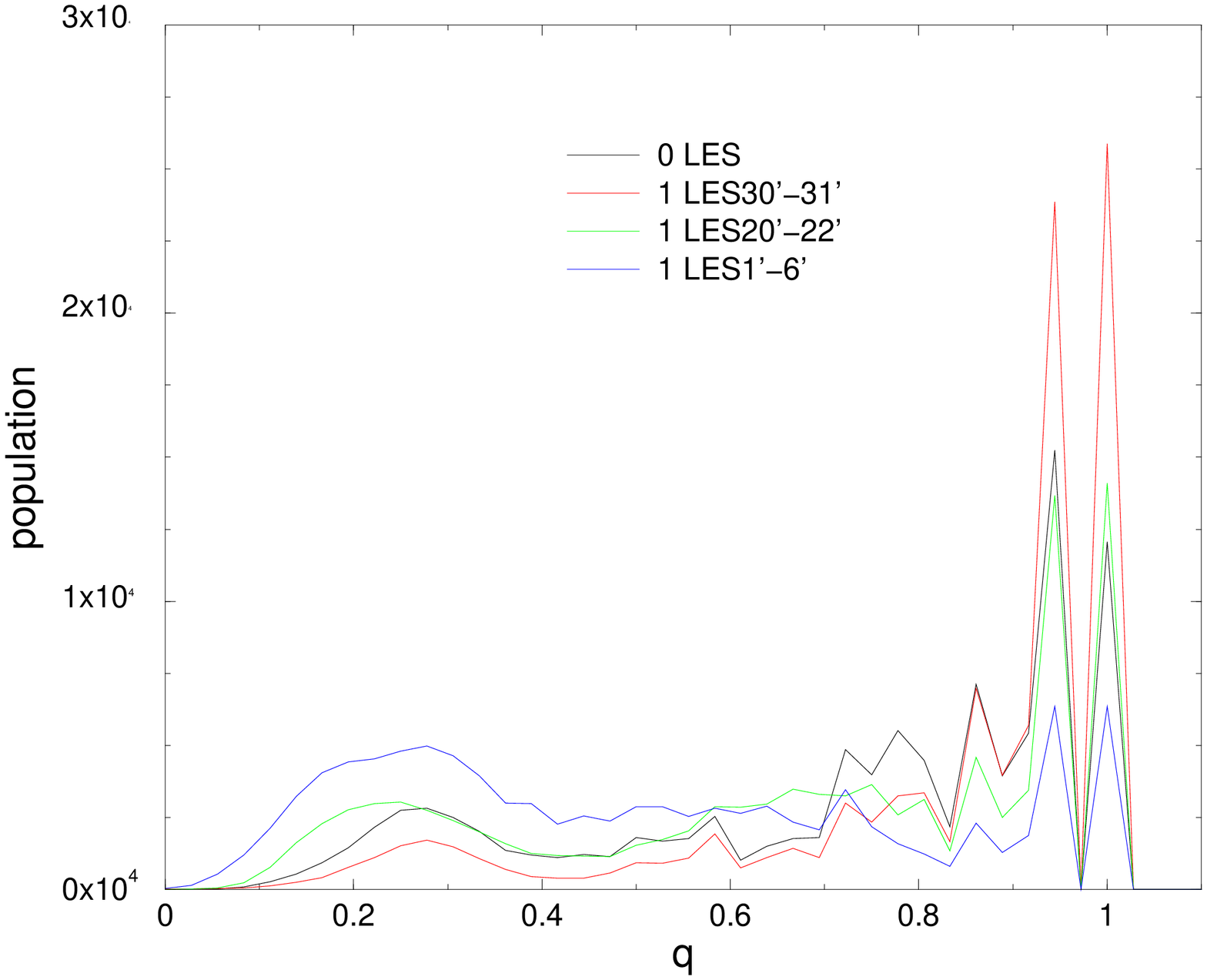,height=6cm,angle=-90}}
\caption{Stability of the protein displayed in Fig. \protect\ref{nonlocal} fodling in presence of $n_{p}=1$ p--LES.}
\label{nonlocal_drug}
\end{figure}

To test the generality of the results discussed in this section we have studied a second 36-mer.  The interest in this second 36-mer is due to the fact that it was designed  minimizing the number of local contacts \cite{sh_prion}. Since local contacts play an important role in the hierarchy of folding events, the behaviour of a protein with as few as possible local contacts is particularly interesting. The primary structure, the LES, the folding nucleus and the native state conformation associated with this 36--mer are shown in Fig. \ref{nonlocal} (a), (b), (c) and (d) respectively.  For this sequence we determined how the presence of one p-LES affects the stability of the protein. The results are shown in Fig. \ref{nonlocal_drug}. The longest p--LES 1'-6' destabilizes the protein (the population of the native state drops from $\sim 50\%$ (without p--LES) to ($29.4\%$) in presence of one p--LES).  On the other hand, the behaviour of the other two p--LES based on the shorter LES (20'--22' and 30'--31', which are "open" LES, according to the definition used in Sect. 5.2) essentially do not affect the stability of the protein. The reason for this behaviour can be understood in terms of specificity. The shorter the p--LES, the higher is the probability that it binds to some part of the protein other than the LES it was designed for. These p--LES, binding not specifically to the protein, are not likely to be effective in denaturing the protein. Since this particular 36--mer protein has been chosen to have few local contacts, it displays some LES which are "marginal", in the sense that lie at the borderline (for their length and for their "open" character) of being able to behave like LES.

\begin{figure}
\centerline{\psfig{file=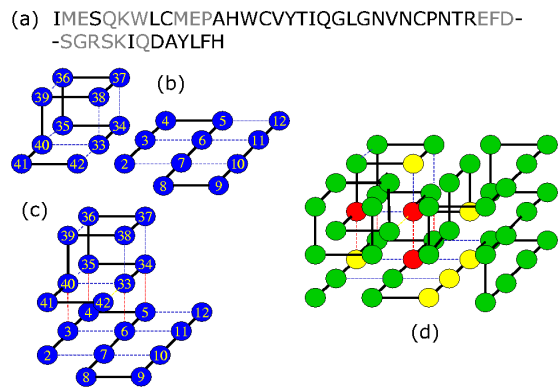,height=6cm,angle=-90}}
\caption{The primary structure of the 48-mer used (a), its LES (b), the folding nucleus (c) and the native state conformation (d).}
\label{48}
\end{figure}

\begin{figure}
\centerline{\psfig{file=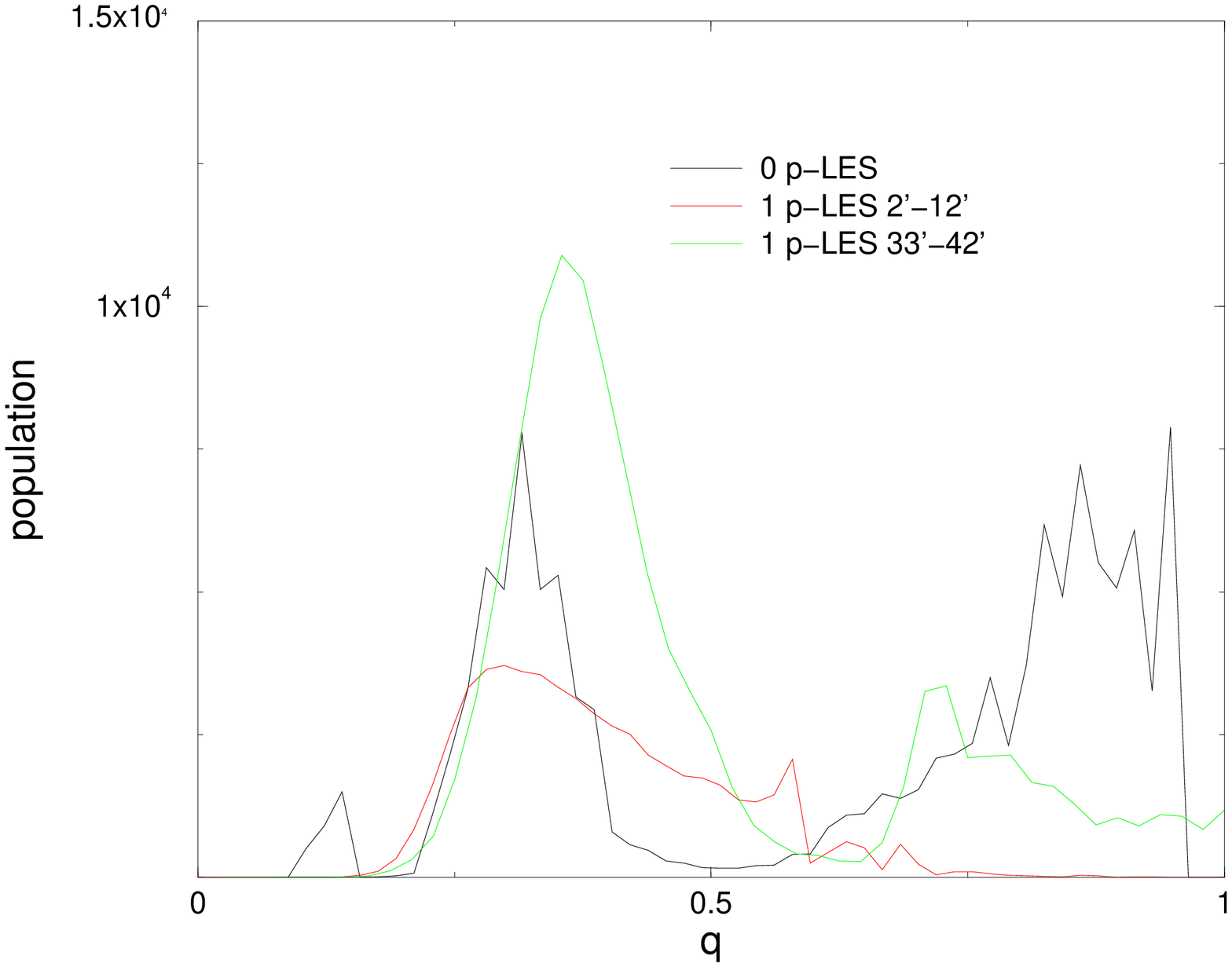,height=6cm,angle=-90}}
\caption{Stability of the protein displayed in Fig. \protect\ref{48} fodling in presence of $n_{p}=1$ p--LES.}
\label{48_drug}
\end{figure}

We have also studied a larger protein, composed of 48 residues. Its primary structure, its LES, folding nucleus and native state conformation are displayed in Figs. \ref{48} (a),(b),(c) and (d), respectively.  For this designed protein we determine how the presence of one p-LES affects the stability of the protein. The results are displayed in Fig. \ref{48_drug} and are in averall agreement with the results obtained in the study of the 36--mers. 

\subsection{Dynamical aspects of the inhibiting mechanism}

\begin{figure}
\centerline{\psfig{file=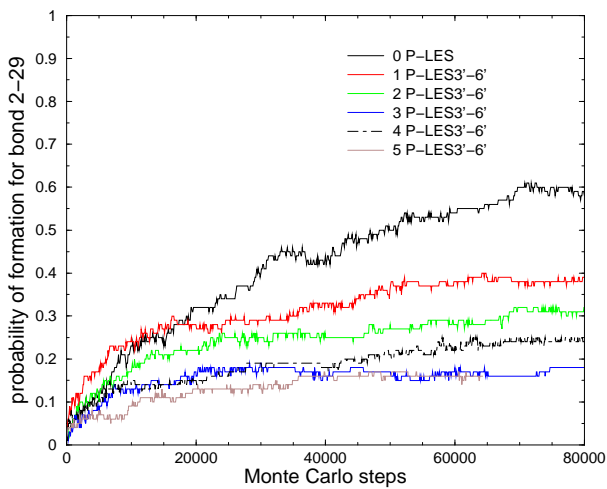,width=8cm,angle=-90}}
\caption{Probability for the bond  3-30 of the protein to be formed during the folding process as a function of the number $n_{p}$(=0,1,2,3,4 and 5) of p-LES 3'--6'.  This contact is taken as representative of the interaction between
\label{drugdyn330}
LES 3--6 and LES 27--30 as a whole.}
\end{figure}

In this section we investigate how the presence of  $n_{p}$ p-LES 3'--6' affects the dynamics of the folding process of the lattice designed protein S$_{36}$. We focus our attention on the dynamics of the formation of contacts 3--6, 3--30 of S$_{36}$, representative of the contacts stabilizing the LES 3--6 and that existing, in the native structure, between the LES 3--6 and LES 27--30. We also study the probability (as a function of the number of MC steps), that a LES interacts with a p-LES. 
In particular, the dynamics of the interaction between site 29 of S$_{36}$ and any amino acid belonging to one of the $n_{p}$ p--LES 3--6' peptides is studied as a function of time (Monte Carlo steps), averaging the contact formation probability over 1000 independent simulations .
One finds  that the formation of bond 3--6, i.e. the formation of LES 3--6 is not affected by the presence of $n_{p}$ p-LES 3'--6' peptides. This testifies to the fact that p--LES do not interfere with the formation of the LES, a phenomenon which occurs very early in the folding process  ($\approx{10^2}$MC steps). On the other hand, the p--LES interferes with the docking of LES that is, with the formation of the folding nucleus, as it is clear from Fig. \ref{drugdyn330}.
Furthermore, the time evolution of the contact between p--LES of type 3'--6' and LES 27--30 of the designed protein is also found to depend on the number of LES.  The higher $n_{p}$, the shorter is the time employed by a p-LES to bind to the corresponding LES.

Summing up, one observes that p--LES interact with LES preventing their docking to form the folding nucleus. Furthermore, simulations indicate that the efficiency of p--LES in denaturing the protein do not depend on its initial conditions (whether the protein is in the native or in an unfolded state).

\subsection{Can the system develop resistance?}

\begin{figure}
\centerline{\psfig{file=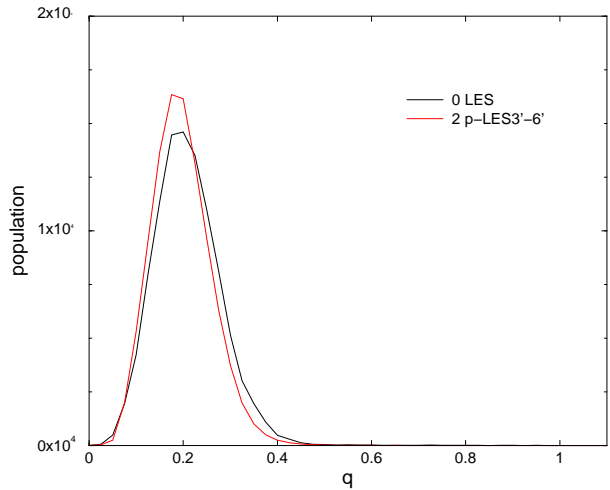,width=8cm,angle=-90}}
\caption{The impact of the point mutation K27G on the folding of S'$_{36}$ where amino acid K which occupies site number 27 has been substituted with amino acid G. The native state is completely destabilized and the activity of the protein is inhibited, in keeping with the fact that site 27 of S$_{36}$ is a hot site (cf. Fig. \protect\ref{1d3d}(d) ). The presence of $n_p=2 $  p--LES3'--6' leaves the population of states unchanged. }
\label{drugres1}
\end{figure}

\begin{figure}
\centerline{\psfig{file=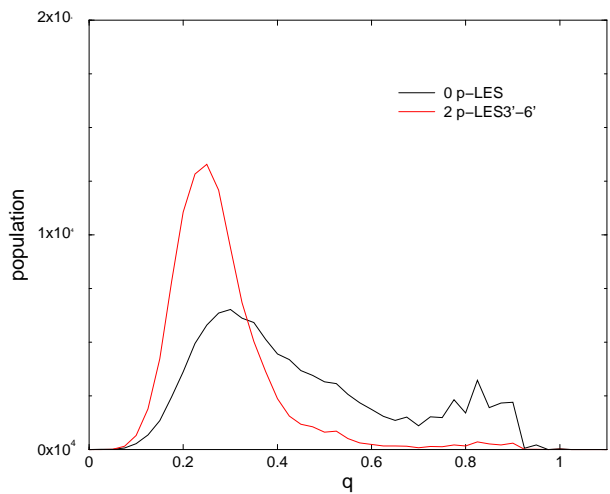,width=8cm,angle=-90}}
\caption{The impact of a point mutation I28S on the folding of S$_{36}$.  In keeping with the fact that site 28 is a cold site (Fig. \protect\ref{1d3d}(d) ) the protein retains its ability to fold although displaying a less stable native conformation (compare the continuous curve of this figure and of Fig \protect\ref{drug3-6}). The drug (p-LES3'-6')is still effective in inhibiting the folding of the mutated protein.}
\label{drugres2}
\end{figure}

One of the main problems related to conventional inhibitor-drugs is the phenomenon of resistance of the lattice designed protein S$_{36}$. Proteins displaying mutations which arise, as a rule, due to the large inaccuracy of genetic regulation associated with retroviruses, that is, viruses which carry their genetic information in a single RNA strand, profit from mutations in the target site of the drug. This is, as a rule, its active site.  Consequently the ability to fold and thus the ability to carry out its biological function is essentially retained, while the drug is no longer able to to bind effectively to the protein. This phenomenon is particularly important, for example, in the case of HIV virus. One of the main proteins involved in the assembly of the virus during its replication,and thus main target of conventional drugs, the HIV-protease, usually develops drug resistance in about 6-8 months from the starting of the therapy (cf. ref. \cite{tomasselli} and refs. therein). Thus, an important question connected with the design of non-conventional  drugs discussed above is: can the system, when targeted by the small peptides p-LES, develop resistance ?.  LES are made up of strongly interacting, as a rule hydrophobic amino acids occupying hot and warm sites well protected inside the protein.  Consequently, one expects that mutations upon the target site of a p-LES (i.e., its complementary LES) lead to a  denaturation or to a conspicuous destabilization of the native state of the protein. To test this expectation the sequence S$_{36}$ was subjected to a drug- induced evolutionary pressure. In other words, simulations of the folding of sequences S$_{36}$', obtained from S$_{36}$ by point mutations, in the presence of $n_{p}$ p-LES were carried out and the results analyzed. Two possible outcomes were found: 1) the mutation leads to a complete denaturation of the protein thus making it totally inactive, 2) the mutated sequence still folds to the native, biologically active state, although the native state is less stable.  In the first case p-LES have no effect on the behaviour of the protein.  In the second case they retain their effectiveness interefering with the folding process and with the stability of the protein, very much as they did in the simulations discussed in Sect. 6.1 and 6.3 (cf. Figs. \ref{drugres1} and \ref{drugres2}).

\subsection{Perspectives on the use of p--LES as drugs}

We have shown how it is possible to inhibit the activity of a protein by blocking its folding with the help of small peptides which mimic the LES.  The very reason why LES make the protein fold fast confers p-LES the features required to a drug to qualify as such: efficiency and specificity. p--LES are efficient because they bind to the protein as strongly as LES bind to each other to form the folding nucleus.  Since LES are responsible for the stability of the protein, their stabilization energy  must be of the order of several times $kT$. These peptides are also as specific as LES are, a specificity which LES have developed in millions of years of evolution to avoid both metastable states and aggregation, let alone make the protein fold fast. The possibility of developing non-conventional drugs in actual situations is tantamount to being able to determine the LES of a given protein.  This, in principle can be done either experimentally, for example, making use of $\varphi$-value analysis (i.e., measuring the relative change in the free energy between the native state and the transition state upon mutation: high $\varphi$-values are associated with "hot" sites \cite{fersht}) or extending the algorithm discussed in Sect. 5.2 with a realistic force field. The resulting peptides can be used directly as drugs, or  as templates to build mimetic molecules, which eventually do not display problems connected with digestion or allergies.  A feature which makes these drugs quite promising as compared to conventional ones is to be found in the fact that the target protein would not be able to evolve through mutations to escape the action of the drug, as it happens, e.g., in the case of viral proteins, because mutations of residues belonging to LES would, in any case, lead to protein denaturation.


\clearpage

\clearpage

\clearpage

\begin{figure}
\centerline{\psfig{file=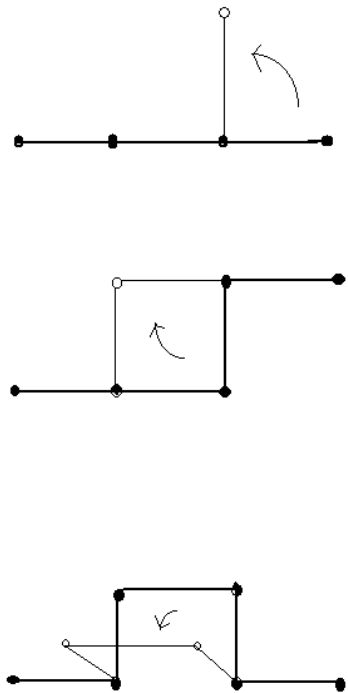,width=8cm,angle=-90}}
\caption{The set of moves used in the Monte Carlo algorithm, known (from top to
bottom) as head/tail move, corner flip and crankschaft, respectively.}
\label{mc_moves}
\end{figure}

\begin{table}
\centerline{\psfig{file=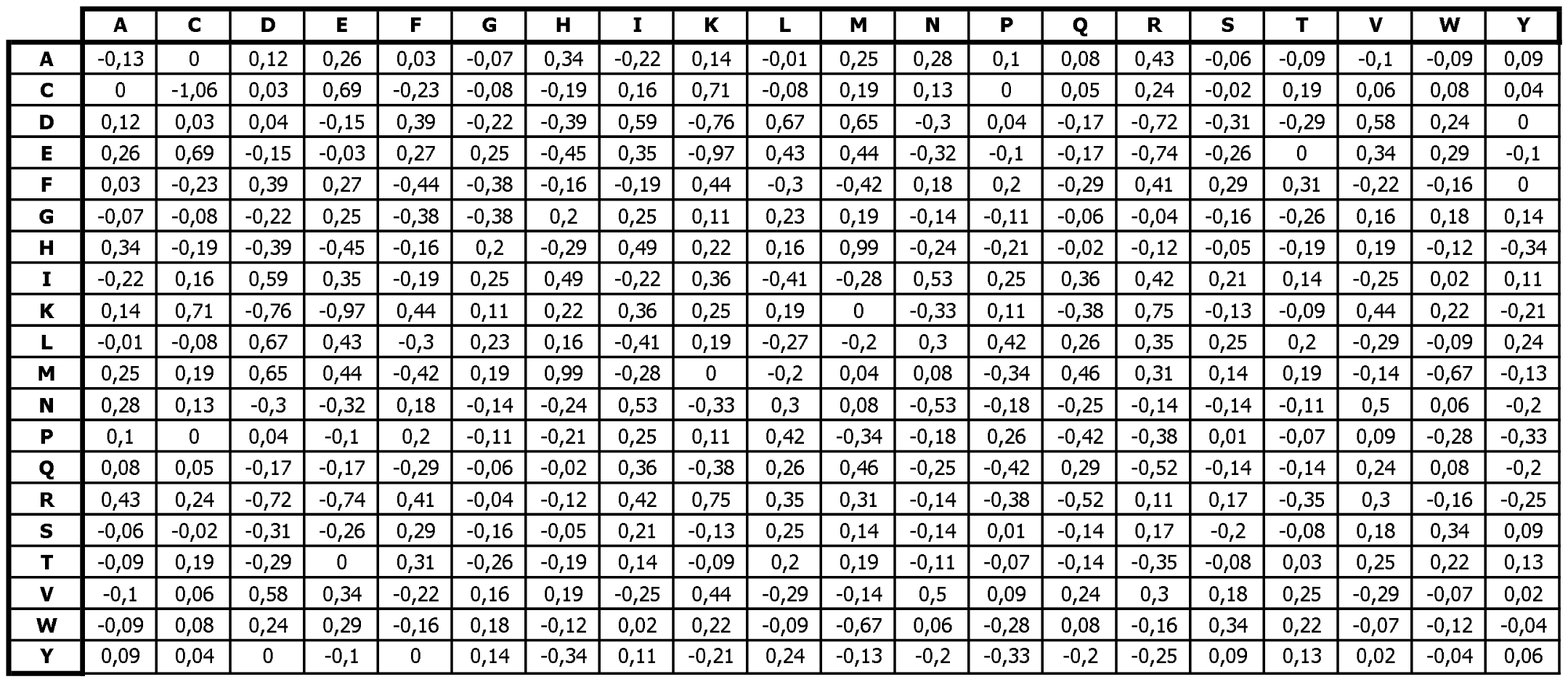,height=18cm,angle=90}}
\caption{The Miyazawa--Jerningan interaction matrix. The average value of the associated matrix elements is equal to $0$, while the standard deviation is $\sigma=0.3$. The matrix elements are expressed in units of $RT_{room}=0.6\;kcal/mol$.}
\end {table}


\appendix
\section{The Monte Carlo algorithm}

The whole thermodynamical information about a protein chain (e.g., the stability of the native state, the folding temperature, etc.) is contained in the partition function
\begin{equation}
\label{part}
Z=\sum_\Gamma \exp(-E(\Gamma)/T),
\end{equation}
where the sum is performed over all the possible conformations of the system and the energy function is, in the present model, that given by Eq. \ref{hamilton}.The huge number of conformations that even a short chain can assume make unfeasible the exact enumeration in Eq. \ref{part} by calculators, nor the function is simple enough to be summed analitically.

The Monte Carlo algorithm \cite{metropolis} is meant to give an estimation of the partition function through the summation of Eq. \ref{part} only over a limited set of conformations. If the choice of this set were made randomly, the algorithm would be rather inefficient (except at high temperatures), since most of the
states display high energy and consequently the associated exponential is small. To solve this problem, the Monte Carlo algorithm builds a Markov chain of states of the system, i.e. a artificial dynamics, which has the purpose of providing
the set over which sum the partition function. The algorithm consists of three steps:
\begin{enumerate}
\item Chosing a random starting conformation of the chain.
\item Performing a random move chosen among a set of permitted elementary moves.\item Accept the move with a probability chosen in such a way that the distribution of states tends, after a large number ov moves, towards a Boltzman distribution.
\end{enumerate}
Steps (2) and (3) are then repeated a large number of times (usually called Monte Carlo steps, or MC steps).

In the present calculations, it is chosen the Metropolis acceptance probability, given by
\begin{equation}
p_{acc}=
\begin{cases}
1& \text{if $\Delta E<0$},\\
\exp(-\Delta E/T)&  \text{ if $\Delta E\geq 0$},
\end{cases}
\end{equation}
where $\Delta E$ is the change in the energy of the chain caused by the (possible) move. The set of moves we used is displayed in Fig. \ref{mc_moves} and is composed of (1) the move of the head or the tail of the chain in a neighbouring site, (2) the flip of a corner conformation and (3) a crankschaft. In principle, this set (as any set of local moves) does not make the system ergodic, a feature which is required by the Monte Carlo algorithm to work, but the subsets of conformational chain which are disjoint from the rest are so tiny, that we can consider effectively the system as ergodic.

Apart from describing the thermodynamics of the system, it has been shown that the Monte Carlo algorithmalso provides a reliable description for the dynamics if the set of moves employed is local \cite{rey}. In fact, it was shown \cite{giapp} that the trajectories resulting from a Monte Carlo simulation with local moves constitute the solution of a diffusive Fokker--Plank equation, which, in turn, is equivalent to Langevin dynamics.


\end{document}